\documentclass[a4paper,fleqn,usenatbib]{mnras}

\newcommand{\lya}{\mbox{${\rm Ly}\alpha$}}

\newcommand{\Av}{$A_V$}
\newcommand{\zqso}{\ensuremath{z_{\textsc{qso}}}}
\newcommand{\zabs}{\ensuremath{z_{\rm abs}}}

\newcommand{\kz}{\ensuremath{\kappa_{\textsc z}}}

\newcommand{\Mgii}{\ion{Mg}{ii}}
\newcommand{\HI}{\ion{H}{i}}
\newcommand{\NHI}{{\rm N_{H\, \textsc{i}}}}
\newcommand{\avgEbv}{\ensuremath{\langle E(B-V) \rangle}}
\newcommand{\logNHI}{$\log({\rm N_{H\, \textsc{i}}})$}
\newcommand{\logNHIcm}{$\log({\rm N_{H\, \textsc{i}}\: /\: cm}^{-2})$}

\usepackage{newtxtext,newtxmath}

\usepackage[T1]{fontenc}
\usepackage{ae,aecompl}

\usepackage{graphicx}	
\usepackage{amsmath}	
\usepackage{amssymb}	
\usepackage{color}
\usepackage{bm}

\title[DLA dust bias in SDSS-II quasar colour selection]{The effect of dust bias on the census of neutral gas and metals in the high-redshift Universe due to SDSS-II quasar colour selection}

\author[J.-K. Krogager et al.]{
	Jens-Kristian Krogager$^{1}$,
	Johan P. U. Fynbo$^{2,3}$,
	Palle M{\o}ller$^{4}$,
	Pasquier Noterdaeme$^{1}$,\newauthor
	Kasper E. Heintz$^{5,3}$
	and Max Pettini$^{6}$
\\
$^{1}$Institut d'Astrophysique de Paris, CNRS-SU, UMR7095, 98bis bd Arago, 75014 Paris, France\\
$^{2}$Dark Cosmology Centre, Niels Bohr Institute, University of Copenhagen, Juliane Maries Vej 30, 2100 Copenhagen \O, Denmark\\
$^{3}$The Cosmic Dawn Center, Niels Bohr Institute, University of Copenhagen, Juliane Maries Vej 30, 2100 Copenhagen \O, Denmark\\
$^{4}$European Southern Observatory, Karl-Schwarzschildstrasse 2, 85748 Garching bei M\"unchen, Germany\\
$^{5}$Centre for Astrophysics and Cosmology, Science Institute, University of Iceland, Dunhagi 5, 107 Reykjav\'ik, Iceland\\
$^{6}$Institute of Astronomy, Kavli Institute for Cosmology, Madingley Road, Cambridge CB3 0HA, United Kingdom
}

\begin{document}

\label{firstpage}
\pagerange{\pageref{firstpage}--\pageref{lastpage}}
\maketitle

\begin{abstract}

	We present a systematic study of the impact of a dust bias on samples of damped Lyman-$\alpha$ absorbers (DLAs). This bias arises as an effect of the magnitude and colour criteria utilized in the Sloan Digital Sky Survey (SDSS) quasar target selection up until data release 7 (DR7).
	The bias has previously been quantified assuming only a contribution from the dust obscuration.
	In this work, we apply the full set of magnitude {\it and} colour criteria used up until SDSS DR7 in order to quantify the full impact of dust biasing against dusty and metal-rich DLAs. We apply the quasar target selection algorithm on a modelled population of intrinsic colours, and by exploring the parameter space consisting of redshift, (\zqso and \zabs), optical extinction, and \HI\ column density, we demonstrate how the selection probability depends on these variables.
	We quantify the dust bias on the following properties derived for DLAs at $z\approx3$: the incidence rate, the mass density of neutral hydrogen and metals, the average metallicity.
	We find that all quantities are significantly affected. When considering all uncertainties, the mass density of neutral hydrogen is underestimated by 10 to 50\%, and the mass density in metals is underestimated by 30 to 200\%.
	Lastly, we find that the bias depends on redshift. At redshift $z=2.2$, the mass density of neutral hydrogen and metals might be underestimated by up to a factor of 2 and 5, respectively. Characterizing such a bias is crucial in order to accurately interpret and model the properties and metallicity evolution of absorption-selected galaxies.

\end{abstract}


\section{Introduction}

	Quasar absorption systems are invaluable tools for our understanding of neutral gas at
	high redshift, as this is currently the only way to study the neutral gas phases beyond
	the local environment where emission studies are sensitive.
	One such class of quasar absorption systems, the so-called damped \lya\ absorbers (DLAs),
	are of particular importance since the high column density of \ion{H}{i}
	($\NHI > 2\times10^{20}$~cm$^{-2}$, \citealt{Wolfe1986}) ensures that the hydrogen is
	predominantly neutral. Due to the low fraction of ionized gas in DLAs, we can infer very
	precise measurements of the abundances of metal species. Moreover, the neutral gas mass
	density in the Universe is dominated by these high column density systems \citep{Lanzetta1995a,
	Prochaska2005, Noterdaeme2009b}. Since the neutral gas phase is needed in order to form
	molecules and subsequently stars, DLAs therefore provide important clues about the gas
	reservoir from which stars form over most of cosmic time.

	Since certain elements (e.g., Fe, Cr, Mn) tend to deplete strongly into dust grains,
	their gas-phase abundances (as measured from absorption lines) are lowered according
	to the strength of the depletion. By studying abundance ratios in DLAs it is therefore
	possible to conclude that some amount of dust must be present in most DLAs
	\citep[e.g.,][]{Ledoux2003, DeCia2016}. Moreover it is found that the optical extinction,
	\Av, scales with $\NHI$ and metallicity \citep{Vladilo2005, Zafar2013}. The more metal-rich
	and high-column density systems will therefore have a higher dust opacity. This will lead
	to a bias against metal-rich and high $\NHI$ systems as these are more likely to be missed
	in optically selected, flux-limited surveys \citep{Pei1991}. The effect of obscuration has
	been studied statistically throughout the last decades, reaching conflicting conclusions
	about the average reddening from DLAs. Most studies agree that the average reddening is low
	but measurements still present a rather large scatter ranging from around 0.002 to 0.01~mag,
	with the strength of the effect varying from weak to no bias in optically selected DLA samples
	\citep[e.g.,][]{Pei1991, Murphy2004, Vladilo2008, Pontzen2009, Frank2010, Kaplan2010, Khare2012}.
	Alternatively, by selecting quasars purely based on their radio properties, the effect of
	a dust bias can be largely overcome as electromagnetic radiation at radio wavelengths does
	not suffer from dust obscuration \citep{Ellison2001c, Jorgenson2006}. Yet, an optical
	counterpart must still be detectable to spectroscopically confirm the source as being
	a quasar. The largest analysis of radio selected quasars has put an upper limit on the
	average reddening from DLAs of $\langle E(B-V) \rangle < 0.040$~mag \citep{Ellison2005}.
	However, the study is limited by the small sample size of only 14 DLAs from a total of 42
	radio-selected quasars with optical and near-infrared data available.
	Combining optical and radio constraints, \citet{Pontzen2009} estimate that a small fraction
	of absorbers (less than around 5\%) will be missed due to dust obscuration, but that the
	mass density of metals, $\Omega_Z$, might be underestimated by up to a factor of two.
	Recently, \citet{Murphy2016} reanalysed the quasar sample of the Sloan Digital Sky Survey
	\citep[][SDSS]{York2000} data release 7 (DR7). The authors find a low average reddening of
	$\langle E(B-V) \rangle = 0.003\pm0.001$~mag and conclude that the optical criteria
	(both in terms of flux and colours) used for target selection in SDSS prior to DR7 have
	a negligible effect on their estimated reddening. While the authors replicate the complex
	selection function of SDSS DR7 to support their conclusion, they do not provide a full
	analysis of the impact of the selection effects on the observed metallicity and $\NHI$ distributions.

	While the number of systems missed due to a dust bias might be small
	(less than 5\% in number, \citealt{Pontzen2009}), these systems are preferentially
	metal-rich and dusty and thus more likely to harbour cold neutral gas and molecules
	\citep{Ledoux2003, Srianand2005, Srianand2008a, Noterdaeme2009a, Noterdaeme2018}.
	It is therefore important to quantify the full implications of dust reddening and
	obscuration for our understanding of the cold ($T\sim100$~K), dust-rich gas phases in DLAs.
	In an attempt to identify this obscured population of dusty DLAs, we have applied
	optical and near-infrared photometric criteria to observe reddened quasars that are
	not targeted by the SDSS selection algorithm \citep{Fynbo2013a, Krogager2015, Krogager2016b}.
	Based on these tailored surveys, we have identified three dusty DLAs at high redshift
	($z \approx 2$) towards quasars that were not flagged as quasar candidates in SDSS
	\citep{Krogager2016a, Fynbo2017, Heintz2018_dustyDLA}. All of these absorbers have
	high metallicity, and while cold ($T\sim100$~K), neutral gas is only securely detected
	in the absorber HAQ2225+0527 \citep{Krogager2016a}, it is highly probable that higher
	resolution spectroscopy of the remaining DLAs will reveal similarly cold, neutral gas.

	In this work, we study the combined effects of dust obscuration and reddening on the
	quasar selection in SDSS DR7 \citep{Richards2002}. In order to assess the selection 
	probability of quasars, we first simulate intrinsic quasar colours and magnitudes as 
	a function of redshift. We then introduce various absorber properties in front of the 
	intrinsic quasar photometry and replicate the SDSS selection algorithm in order to 
	quantify the effect of various parameters. Following the analysis of \citet{Pontzen2009}, 
	we use our calculated selection probabilities to infer the intrinsic distribution 
	functions for metallicity and $\NHI$, and find that the full, combined optical selection 
	(using both colour and flux-limit) leads to a significantly stronger bias than assuming 
	a flux-limit alone. This is somewhat alleviated if assuming that a small fraction of 
	quasars are selected irrespective of their optical properties (similar to the cross-matching 
	to radio sources performed in the SDSS).
	We stress that all references to SDSS throughout this work refers solely to the first 
	epochs of SDSS (I and II up until data release 7) before the onset of the Baryon Oscillation 
	Spectroscopic Survey (BOSS, also referred to as SDSS-III), since the spectroscopic target 
	selection changed completely from the algorithm described in \citet{Richards2002} for 
	SDSS I/II to the more complex algorithms used for BOSS 
	\citep{Richards2009, Yeche2010, Kirkpatrick2011, Bovy2011, Ross2012}.
	While the spectroscopic target selection for quasars is more complex for the BOSS sample,
	the selection is still based on optical information and is therefore still susceptible to
	biases.
	Moreover, the average reddening effect by DLAs has only been studied for DLAs selected
	from the quasar sample of SDSS data-release 7 (DR7). We therefore do not consider the
	quasar samples of later data-releases in the current work. The effect of the BOSS quasar
	selection will be studied in a forthcoming paper.

	This paper is organized as follows: In Sect.~\ref{simulation}, we describe our
	simulated photometry and the calculation of selection probabilities; In Sect.~\ref{dustbias},
	we present our calculation of fractional completeness of absorption-derived quantities
	such as the gas mass density of neutral gas and metals in DLAs; The systematic uncertainties
	related to our modelling are investigated in Sect.~\ref{systematics}; In Sect.~\ref{discussion},
	we discuss the results and implications of our work as well as the related caveats,
	and lastly in Sect.~\ref{summary}, we summarize our results.

	Throughout this work, we will assume a flat $\Lambda$CDM cosmology with
	$H_0=68\, \mathrm{km s}^{-1}\mathrm{Mpc}^{-1}$, $\Omega_{\Lambda}=0.69$
	and $\Omega_{\mathrm{M}} = 0.31$ (Planck Collaboration 2014)\nocite{Planck2014}.

\section{Quasar Colour Selection}
\label{simulation}

	We investigate the effect of dust reddening by replicating the colour and magnitude
	selection algorithm of SDSS. As input to the selection algorithm, we simulate an
	intrinsic population of quasars at different redshifts (\zqso) with varying amounts
	of foreground dust (\Av) at various absorption redshifts (\zabs) for two broad classes
	of extinction laws (SMC and LMC; as quantified by \citealt{Gordon2003}).
	Moreover, we include the reddening effect of strong \ion{H}{i} absorption in our model.
	In the following section, we will describe the details of how we simulate the intrinsic
	quasar properties, how we apply the SDSS quasar selection algorithm, and how we simulate
	the absorption properties and calculate the selection probabilities.
	Using a physically motivated model in which metallicity ($Z$) and $\NHI$ scale with \Av,
	we then calculate the selection probabilities as a function of $Z$ and $\NHI$.
	We use this 2-dimensional selection probability to fit intrinsic
	distributions of $Z$ and $\NHI$ in order to reproduce the observed distributions
	of $Z$ by \citet{Rafelski2012} and \citet{Jorgenson2013} and of $\NHI$ by
	\citet{Noterdaeme2009b} as well as the observed average reddening measured by
	\citet{Murphy2016}.
	Lastly, we calculate the fractional completeness in various observable quantities
	for DLAs following \citet{Pontzen2009}.

\subsection{Constructing the Statistical Samples}
\label{stat_samples}
	
	In order to make a meaningful statistical comparison between our model and the data,
	we only consider quasars brighter than the magnitude-limit used for the complete sample
	of quasars in SDSS ($i \leq 19.1$). The quasar sample of SDSS does include a significant
	number of targets fainter than this limit, but the completeness is low
	($\sim30$~\% for \zqso~$\approx3$ based on the luminosity function at this redshift).
	It is therefore not possible for us to replicate this deeper selection
	(referred to as `high-$z$ quasar selection' or `$griz$'-selection,
	see Sect.~\ref{quasar_selection}) in a statistically meaningful way.

	Since the data by \citet{Rafelski2012}, \citet{Noterdaeme2009b} and \citet{Murphy2016} are
	based on the full sample of DLAs in SDSS up until data-release 7 (DR7) irrespective of the
	brightness of the background quasar, we first have to clean the samples and remove DLAs observed
	towards quasars that do not meet the $i$-band flux-limit. The sample by \citet{Rafelski2012}
	is composed of a purely SDSS-selected sub-sample from DR5 (their table~2) together with a sample
	compiled from the literature (their table~3). Since a large part of the literature sample
	is not covered by the SDSS foot-print, we are not able to obtain SDSS $i$-band magnitudes.
	Instead, we have used available $i$-band photometry from the Pan-STARRS1 \citep{Chambers2016}
	as this survey uses a very similar $i$-band to that of SDSS. When no photometry is available
	from neither SDSS nor Pan-STARRS1, we have used available broad-band photometry in $B$, $V$,
	and $R$ bands to infer an $i$-band magnitude based on colour transformations between the
	various magnitude systems. In those cases, the targets are several magnitudes brighter than
	the $i\leq19.1$ limit, and the large uncertainty in the colour transformations is therefore
	not important.
	For one target (J0242$-$2917) we have not been able to obtain any photometric information. This
	object has been excluded from the final sample.
	Based on the photometry, we discard all DLAs for which the background quasar is fainter than
	$i=19.1$.
	Moreover, we exclude targets where the metallicity is based on a combination of limits or
	on refractory elements alone (Fe, Ni, Cr, etc.).
	
	We furthermore include the DLA sample of metallicities by \citet{Jorgenson2013}
	homogeneously selected from SDSS DR5 with $i<19$. However, we remove duplicates between
	the two samples giving preference to measurements based on sulphur over those based on
	silicon and similarly we give preference to higher resolution measurements.
	As for the `Rafelski sample', we remove measurements based on refractory elements or limits.
	This results in a total sample of 202 DLAs with good quality metallicity measurements.
	
	We highlight that the quasar selection algorithm did not change up until DR7 and hence
	all SDSS data-releases before and including DR7 are based on the selection algorithm
	as presented by \citet{Richards2002}.
	The literature sample included by \citet{Rafelski2012} is not strictly selected
	based on the same criteria as the SDSS selection algorithm; however, since quasars before
	SDSS were mainly identified either via radio observations or the UV excess method, both of
	which are explicitly included in the SDSS algorithm, we argue that the
	literature sample of \citeauthor{Rafelski2012} would very likely have been targeted
	by the SDSS selection algorithm that we study in this work.
	This claim is further bolstered by the fact that such `pre-SDSS' quasars were incorporated
	in the sample of quasars used to optimize the SDSS algorithm \citep{Richards2002}.

	The samples by \citet{Noterdaeme2009b} and \citet{Murphy2016} are derived purely
	from SDSS-DR7 and we can therefore easily clean out the DLAs in front of quasars
	fainter than $i=19.1$. The average reddening measurement by \citet{Murphy2016}
	is $\langle E(B-V) \rangle = 3 \pm 1$~mmag.
	We find a consistent average reddening for the flux-limited sub-sample yet with
	a slightly larger uncertainty: $\langle E(B-V) \rangle = 3.0 \pm 1.5$~mmag.

	Similarly, we find no effect on the $\NHI$ distribution function derived by
	\citet{Noterdaeme2009b} when restricting the sample to quasars with $i<19.1$.

\subsection{Simulating Quasar Photometry}
\label{simulation_quasars}

\begin{figure}
	\centering
	\includegraphics[width=0.48\textwidth]{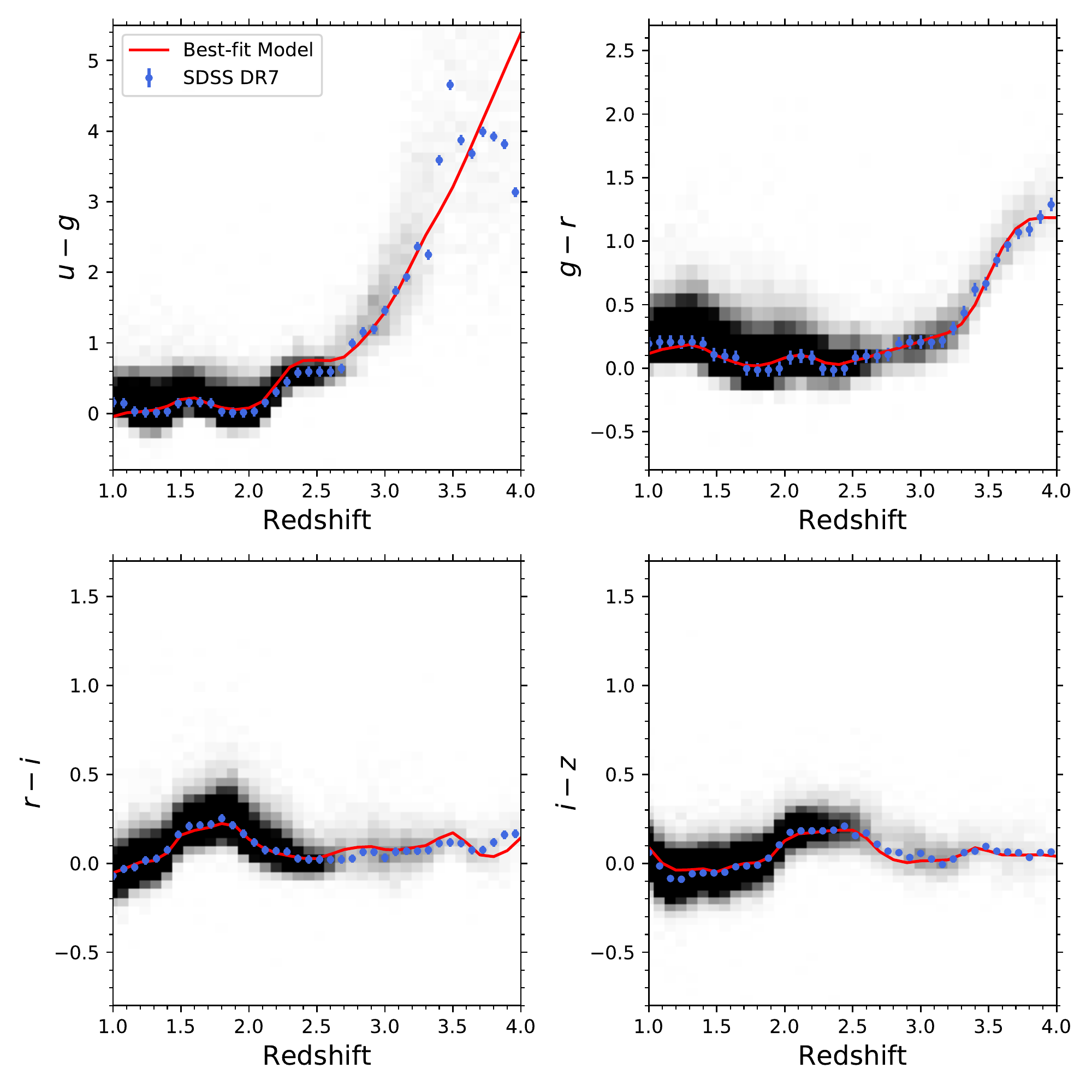}
	\caption{Colour--redshift relations for quasars in SDSS DR7.
	The underlying gray-scale distribution indicates the number of quasars.
	The mode of the observed distribution as a function of redshift
	is shown as blue dots with errorbars corresponding to half the bin-size.
	The red curve shows the best-fit quasar model for
	$\Delta\alpha = -0.01\pm0.06$ and $w_{\textsc{igm}} = 0.67\pm0.01$.}
	\label{fig:col_redshift}
\end{figure}

\begin{figure}
	\centering
	\includegraphics[width=0.48\textwidth]{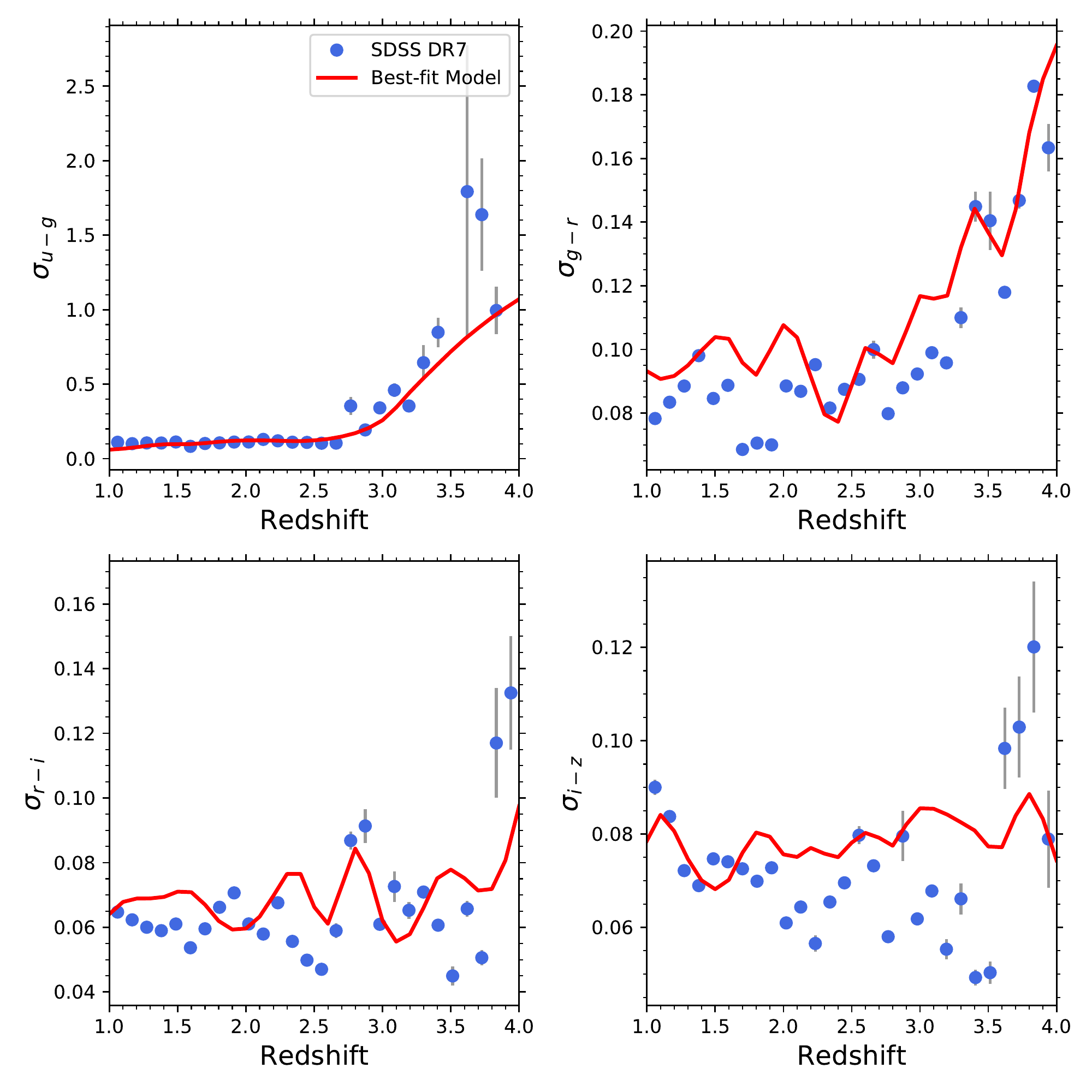}
	\caption{Width of colour--redshift relations for quasars in SDSS DR7.
	The red curve shows the best-fit quasar model for
	$w_{\alpha} = 1.5\pm0.2$ and $\sigma_{\textsc{igm}} = 0.13\pm0.01$.}
	\label{fig:width_redshift}
\end{figure}

	In order to generate proper input for the SDSS quasar selection algorithm, we need
	not only the colour information but also the magnitudes in the 5 filters of SDSS
	($u$, $g$, $r$, $i$, and $z$). We therefore draw samples from the quasar luminosity
	function as a function of quasar redshift \citep{Manti2017} down to a limiting
	magnitude of $i<19.1$. To get the right distribution of colours we construct a
	fiducial quasar model based on available constraints from the literature.
	For this purpose, we use the quasar
	template by \citet{Selsing2016} as a basis for the spectral energy distribution;
	however, as the template only covers rest-frame wavelengths down to $\sim$1000~\AA,
	we extend the template bluewards of $\lambda_{\rm rest}=1000$~\AA\ by assuming
	a power-law with a spectral index of $\alpha_{\lambda}=-0.3$ \citep{Lusso2015}.
	This break wavelength corresponds fairly well to the break in the power-law
	spectral shape observed by \citet{Lusso2015}, who infer $\lambda_{\rm break} \sim 912$~\AA.
	The template is redshifted to the given \zqso\ and scaled to the absolute flux as
	determined from the luminosity function.
	This results in a 5-element vector of magnitudes per band denoted $\vec{m}_{0,i}$
	referring to the $i$-th random realisation of quasar photometry drawn from the
	luminosity function.
	
	Since the intrinsic properties of quasars are known to vary from object to object,
	we include variations in the intrinsic power-law spectrum. The change in power-law index
	is applied as a relative offset ($\Delta\alpha$) with respect to the intrinsic value
	of $\alpha_{\lambda}=-1.70$ \citep{Selsing2016}. For the fiducial model, $\Delta\alpha$
	has a value of $0$.
	Following \citet{Krawczyk2015}, we assume that the power-law index is distributed as a Gaussian
	centred at the intrinsic value of the template with a width of $\sigma_{\alpha} = 0.186\pm0.001$.
	Rather than randomly assigning a value of the power-law index for each realisation,
	we calculate the offset in magnitudes per band as a function of redshift relative to the
	intrinsic template for a change of $\sigma_{\alpha}$ in power-law index.
	This allows us to evaluate the mean colors at a given redshfit without having to evaluate
	a large ensemble.
	The pre-evaluated offsets per band is denoted as the 5-element vector, $\vec{\Delta}_{\alpha}(z)$.
	For each quasar realisation, we then draw a random number from a Normal distribution
	$X_i \sim \mathcal{N}(\mu=0; \sigma=1)$ and multiply $\vec{\Delta}_{\alpha}(z)$ by $X_i$.
	Lastly, in order to take into account variations of emission lines, we scale
	$\vec{\Delta}_{\alpha}(z)$ by the parameter $w_{\alpha}$, which for the fiducial model
	has the value $w_{\alpha} = 1$.
	
	The last effect we include in the fiducial model is the attenuation bluewards of
	the quasar \lya\ emission line due to hydrogen Lyman-series absorption from the
	intergalactic medium.
	We use the theoretically derived average attenuation for different redshifts as a
	function of wavelength by \citet{Meiksin2006} to calculate the redshift-dependent
	average attenuation per band denoted $\vec{A}_{\textsc{igm}}(z)$\footnote{The average
	attenuation is calculated by interpolating between the values given by \citet{Meiksin2006}.}.
	The attenuation is assigned a randomly drawn weight from a Normal distribution:
	$Y_i \sim \mathcal{N}(w_{\textsc{igm}}, \sigma_{\textsc{igm}})$,
	where $w_{\textsc{igm}}$ is an overall normalization of the theoretical prediction
	in order to match the data, and $\sigma_{\textsc{igm}}$ determines the dispersion in
	IGM absorption from one object to another.
	This mimics effects of random sight lines through the \lya\ forest.
	For the fiducial model, the parameters of this distribution are:
	$(w_{\textsc{igm}}, \sigma_{\textsc{igm}}) = (1.0, 0.2)$.
	
	The final 5-band photometry for a given realisation, $\vec{m}_i$, is then calculated
	as:
	\begin{equation}
		\label{eq:qso_model}
		\vec{m}_i = \vec{m}_{0,i}(\Delta\alpha, z)
		+ w_{\alpha} X_i \vec{\Delta}_{\alpha}(z)
		+ Y_i(w_{\textsc{igm}}, \sigma_{\textsc{igm}}) \vec{A}_{\textsc{igm}}(z) ~.
	\end{equation}

	For each of the realizations of quasar magnitudes in the 5 SDSS filters, we assign
	photometric uncertainties following an empirically derived model between magnitude
	and uncertainty inferred from point source data in SDSS DR7. We fit the distribution
	of photometric uncertainties as a function of observed magnitudes for each of the
	five bands using a 4th order polynomial. For sources with magnitudes larger than
	$22.5$~mag, we assign a constant photometric uncertainty. This does not significantly
	affect our simulations, as targets this faint will not meet the $i$-band criterion
	and hence will not be considered for target selection.
	Following \citet{Richards2002}, we increase the photometric uncertainties by 0.0075~mag
	(added in quadrature) to account for systematic uncertainties in the flux calibration.
	The authors moreover add an error contribution from the correction for Galactic extinction,
	for which they add in quadrature 15\% of the extinction value to the photometric uncertainty.
	Since we do not simulate the projected sky position of our targets, we simply assume
	a median correction in the five filters of $A_{\rm med}=0.184, 0.135, 0.098, 0.074, 0.053$ for
	$u$, $g$, $r$, $i$, $z$, respectively. The median Galactic correction values are derived
	from 10\,000 quasars in SDSS DR7. We then add in quadrature 15\% of the median for each
	filter to the uncertainty of the given filter.

	In order to optimize this fiducial model, we fit the 4 colour relations of the SDSS
	colour-space ($u-g$, $g-r$, $r-i$, $i-z$) as a function of redshift. We parametrize
	the colour--redshift relations by measuring the `mode' and standard deviation of the
	4 colour distributions in redshift bins of $\Delta z\approx0.1$. In order to reproduce
	the mode of the observed colour--redshift distribution, we only need to vary the
	parameters: $\Delta\alpha$ and $w_{\textsc{igm}}$.
	The widths of the distributions are not fitted during the same process, since they do
	not alter the mode of the distributions.
	We will instead constrain the width of the distributions in a subsequent step.
	We note that we fit the mode of the observed distributions since this quantity will
	be less affected by dust reddening compared to the mean and the median of the colour
	distributions. The best-fit model parameters are
	$\Delta\alpha = -0.01\pm0.06$ and $w_{\textsc{igm}} = 0.67\pm0.01$;
	However, we find that in order to fit the $u-g$ colour properly we need to increase
	the IGM absorption in the $u$-band by a factor $w^u_{\textsc{igm}} = 1.5$.
	The resulting colour--redshift relations are shown in Fig.~\ref{fig:col_redshift}.
	
	In order to constrain the parameters $w_{\alpha}$ and $\sigma_{\textsc{igm}}$,
	we fit the width of the colour--redshift distributions while keeping the parameters
	$\Delta\alpha$ and $w_{\textsc{igm}}$ fixed to their best-fit values.
	In Fig.~\ref{fig:width_redshift}, we show the model for the best-fit parameters
	$w_{\alpha} = 1.5\pm0.2$ and $\sigma_{\textsc{igm}} = 0.13\pm0.01$.
	Although we include uncertainties in Fig.~\ref{fig:width_redshift}, these are not
	taken into account in the fit, since this would give much stronger weight to
	the low-redshift bins where there are many more quasars. In order to give equal
	weight to all redshifts during the fit, we use an average uncertainty of 0.01
	for all measurements.

	The intrinsic properties of the simulated data match well the observed properties
	of spectroscopically confirmed quasars as shown in Figs.~\ref{fig:col_redshift} and
	\ref{fig:width_redshift}. This allows us to probe the overall redshift dependence
	of the quasar selection algorithm (see Sect.~\ref{quasar_selection}).
	Yet, there are still significant discrepancies for certain redshift intervals.
	While it is beyond the scope of the current work to reproduce
	in detail the quasar properties over the full redshift range, we do attempt to
	optimize the model in certain redshift ranges of interest. Specifically for quasars
	at $\zqso = 3$ and $2.5$, which we use for the detailed analysis of dust bias against
	absorption systems in Sect.~\ref{dustbias}, we optimize the quasar model.
	The additional optimization and robust evaluation of model uncertainties are
	described in Appendix~\ref{app:opt}.

	In the quasar model described above, we have neglected the contribution of BAL quasars
	as these are anyways not included in the sample of quasars used for absorption analyses.
	Other possible effects that could affect our intrinsic quasar photometry model would be
	systematic variations in the emission lines, e.g., the Baldwin effect \citep{Baldwin1977},
	blazars with no apparent emission lines, or quasars with abnormal spectral shapes
	\citep{Meusinger2012}. However, as blazars and quasars with abnormal spectral shapes
	contribute very little to the overall population of quasars \citep{Meusinger2012, Li2015},
	we have neglected the contribution of these in order to keep the model as simple as
	possible.

\subsection{SDSS quasar candidate selection}
\label{quasar_selection}

	The simulated photometry and errors described above are then passed through the
	steps of the selection algorithm as outlined in figure~1 of \citet{Richards2002}.
	The calculation is split into two selection `branches':
	$ugri$, selected in the 3-dimensional $u-g$, $g-r$, and $r-i$ colour space,
	and $griz$, selected in the 3-dimensional $g-r$, $r-i$, $i-z$ colour space.
	We have re-implemented the definition of the `stellar locus' in $ugri$ and $griz$
	colour-spaces by \citet{Richards2002}. The details of our new implementation in
	Python are given in Appendix~\ref{app:algorithm}. Since we are working on simulated
	photometry, we do not implement the checks for photometric data quality flags.
	Moreover, these photometric impurities will not introduce any significant bias
	but merely lower the completeness.
	Lastly, we apply the same magnitude cut as \citet{Richards2002}, namely, 
	$i< 19.1$~mag and $i< 20.2$~mag for targets in the $ugri$ and $griz$
	colour-spaces, respectively.
	However, for our statistical analysis, we use a single magnitude cut of $i< 19.1$.
	The resulting fraction of simulated quasars identified through the algorithm will
	hereafter be referred to as the selection probability, $P$.

	Since we only consider quasars brighter than $i<19.1$ for the statistical
	bias calculation in this work, the total selection probability is then given
	by the union of the two selection branches, and simultaneously fulfilling the
	$i$-band criterion:
	\begin{equation}
		P_{\rm tot} = \frac{N(ugri \cup griz\ \&\ i < 19.1)}{N(i < 19.1)}
	\end{equation}

	\noindent
	where $N(ugri \cup griz)$ denotes the number of quasars that are selected
	either by the $ugri$ or $griz$ criteria.

\begin{figure}
	\centering
	\includegraphics[width=0.95\columnwidth]{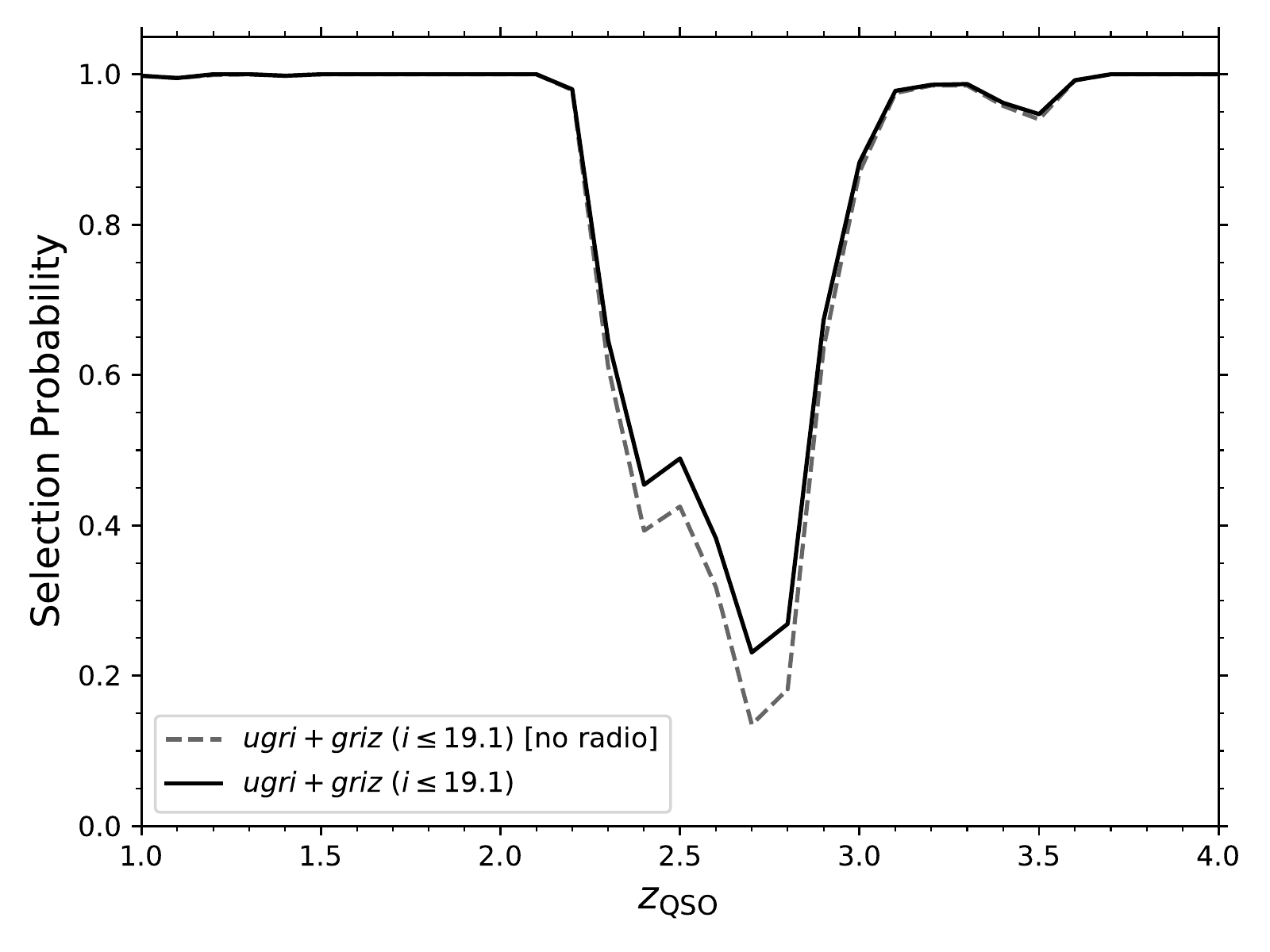}
	\caption{Selection probability as function of \zqso\ for 1000 simulated
		quasars in each redshift bin.
		The solid line shows the full selection probability
		($ugri$ + $griz$) with radio selection for quasars brighter
		than $i\leq 19.1$. The gray, dashed line shows the
		effect of turning off the radio selection.}
	\label{fig:intrinsic_targets}
\end{figure}

	Lastly, we implement a mock radio selection corresponding to the cross-matching
	between FIRST \citep{FIRST} and SDSS by simply allowing a constant fraction of quasars
	to be selected irrespective of their optical colours. Here we assume a fraction of
	radio detections of 10\% inferred from FIRST cross-matching in independent surveys
	\citep{Hewett2001, Krogager2016b}. Thus, 10\% of the simulated quasars will be allowed
	to pass the quasar selection irrespective of the colour criteria;
	Nevertheless, the quasars still have to pass the $i$-band magnitude limit.
	The result of passing the intrinsic quasar photometry as described in
	Sect.~\ref{simulation_quasars} is shown in Fig.~\ref{fig:intrinsic_targets}.
	We recover the strong drop in selection probability for quasars with redshift
	$z\sim2.5$ as demonstrated in the original work by \citet{Richards2002}.
	This low sensitivity for quasars in this redshift range is caused by the fact
	that the quasar colours cross the stellar locus in the optical colour spaces
	utilized by the SDSS.
	For the redshift range $2.3 \leq \zqso \leq 3.0$, we have carried out a similar
	optimization of the quasar model as described in Appendix~\ref{app:opt}.

\begin{figure*}
	\centering
	\includegraphics[width=0.95\textwidth]{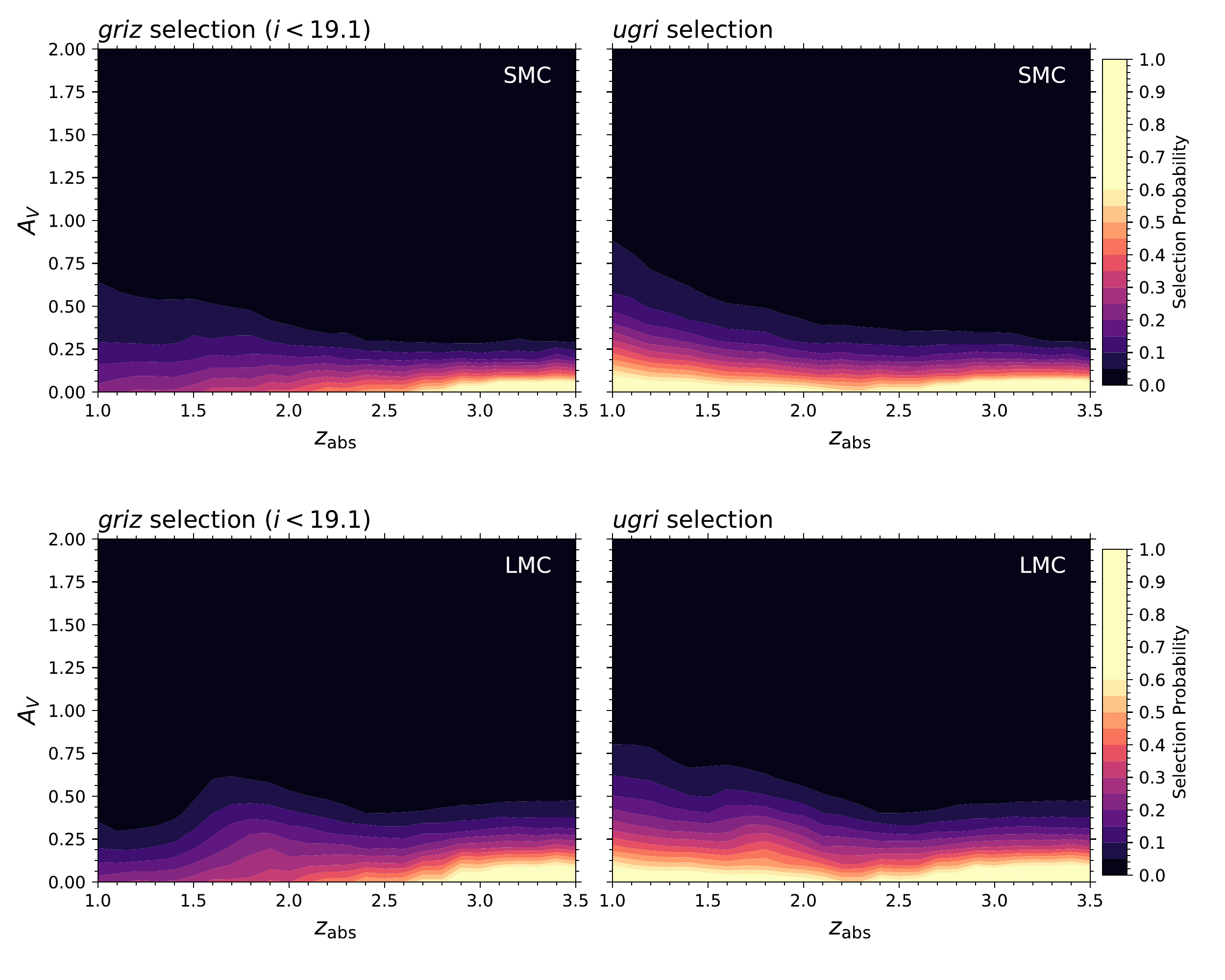}
	\includegraphics[width=0.57\textwidth]{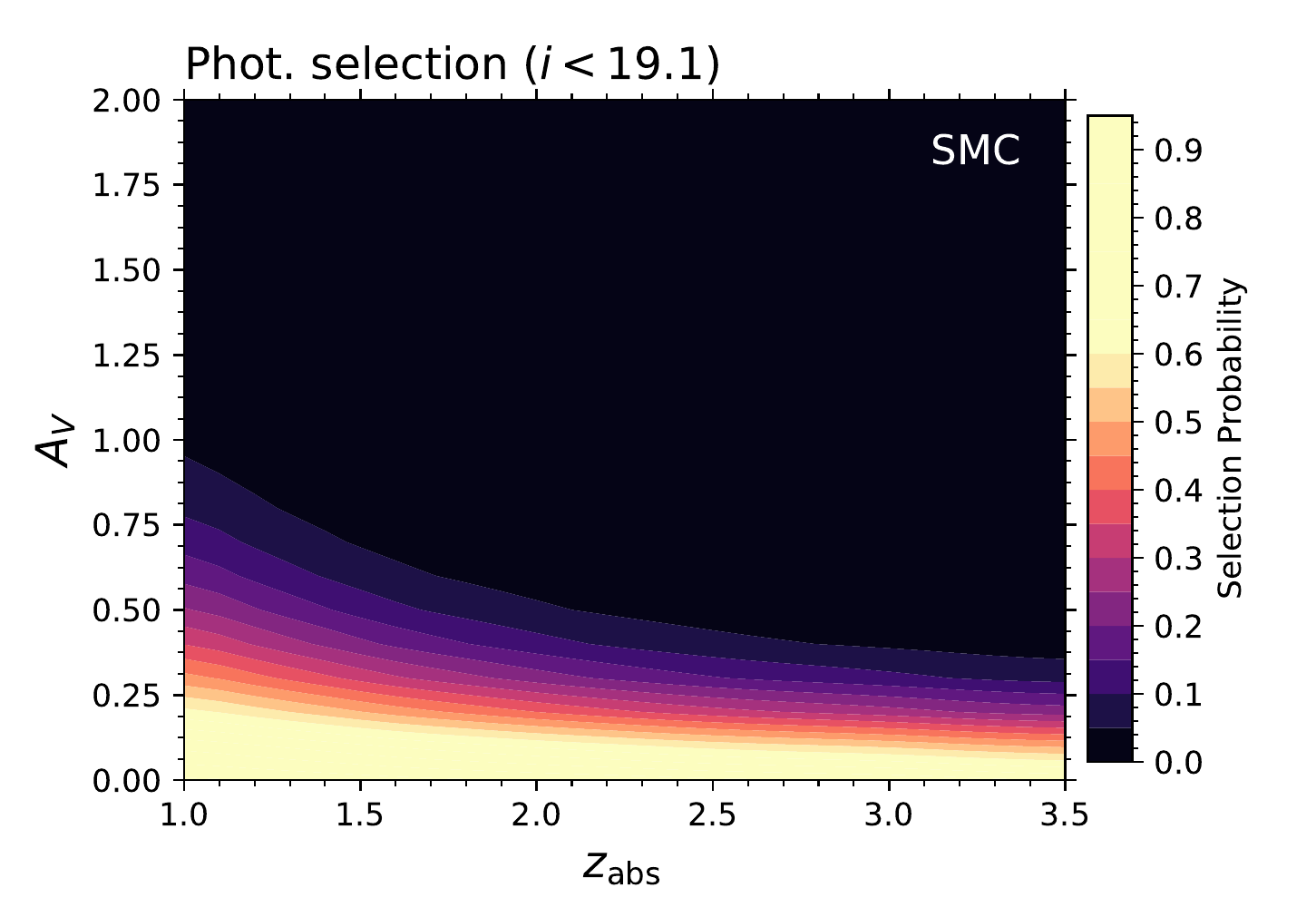}
	\caption{Selection probability for the $ugri$ and $griz$ selections as function of absorber
		redshift and \Av. The colour coding represents the selection probability based on
		the $griz$ (only for $i<19.1$; left panels) and $ugri$ (right panels) selection criteria.
		The two upper rows show the probabilities for the colour and magnitude
		selection for SMC- and LMC-type extinction curves, respectively.
		The bottom row shows the probability for a purely flux-limited selection
		assuming SMC-type extinction.}
	\label{fig:Av_zabs}
\end{figure*}

\subsection{Absorption Properties}
\label{simulation_abs}

	We here investigate the effect of applying intervening dust reddening and strong
	\ion{H}{i} absorption to the quasar photometry. In the first part, we will simply
	calculate the selection probability for the various parameters in order to visualize
	the dependence of the selection probability on these variables. We therefore sample
	all properties uniformly and calculate the selection probability in this parameter
	space.
	In the second part, we will use a physically motivated model to calculate how the
	selection probability affects the observed distributions of metallicity and column
	density of \ion{H}{i}.

	When simulating the absorber properties, we set up a grid of quasar redshifts in the
	interval $1< \zqso <3.6$ (in steps of 0.2), with absorber redshifts in the range
	$1.0< \zabs <3.5$ in steps of 0.1 and requiring that $\zabs \leq \zqso$,
	with \Av\ in the interval $0<$~\Av~$<2$~mag in steps of 0.1, and with \logNHI\ in
	the range $20<$~\logNHIcm~$<23$ in steps of 0.3.
	For each point in this 4-dimensional parameter space (\zqso, \zabs, \Av, \logNHI),
	we then simulate an ensemble of 1000 quasars as described in Sect.~\ref{simulation_quasars}
	and apply the properties \Av\ and \logNHI\ to the simulated quasar photometry.
	The ensemble is then passed through the SDSS quasar selection algorithm and the
	selection probability is calculated. We perform this calculation twice assuming
	different reddening laws: either SMC- or LMC-type as parametrized by \citet{Gordon2003}.

	In Fig.~\ref{fig:Av_zabs}, we show the resulting $P_{ugri}$ and $P_{griz}(i<19.1)$
	in the (\zabs, \Av)-grid averaged over all quasar redshifts and all values of \logNHI.
	For comparison, we show the selection probabilities of quasars for a purely flux
	limited selection of $i<19.1$. These flux-limited-only calculations were performed
	on the same simulated data simply by skipping the colour selection criteria and only
	applying the $i$-band magnitude cut.
	In Appendix~\ref{app:figures}, we show the averaged $P_{ugri}$ and $P_{griz}(i<19.1)$
	in the (\zabs, \zqso) and (\Av, \zqso) parameter spaces respectively.\\

	In the above analysis, we have not assumed any physical relation between absorber
	properties and \Av, however, it is well known that \Av\ scales with both \logNHI\
	and metallicity \citep[e.g.,]{Vladilo2005, Zafar2013}. We therefore run a similar
	set of models as described above assuming a fixed dust-to-metals ratio
	($\kappa_{\rm z}$) to investigate the effect of the colour selection on the
	$Z$ and \logNHI\ distribution of absorbers.
	We use the relation by \citet{Zafar2013} to quantify the $V$-band dust extinction
	in terms of $Z$ and \logNHI:
	\begin{equation}
	\label{eq:dust}
		\log(A_V) = \log({\rm N_{H\,\textsc{i}}}) + \log(Z/Z_{\odot}) + \log(\kz),
	\end{equation}
	
	\noindent
	where $\kz$ is the dust-to-metal ratio.
	
	In order to compare our results to the previous work by \citet{Pontzen2009}, we will
	assume an absorption redshift of \zabs~$=3.0$ which corresponds to the median redshift
	of absorbers in the analysis by \citet{Pontzen2009}.
	While \citet{Pontzen2009} do not worry about the redshift of the background quasar,
	we must take this into account when estimating the bias of the full SDSS selection
	criteria, since the colours depend on \zqso. In the following, we assume that the
	background quasars are at an average \zqso~$=3.2$. We will also investigate the
	selection probability at lower redshift ($z\approx 2$) to demonstrate the redshift
	dependence of the bias.

\begin{figure}
		\centering
		\includegraphics[width=0.5\textwidth]{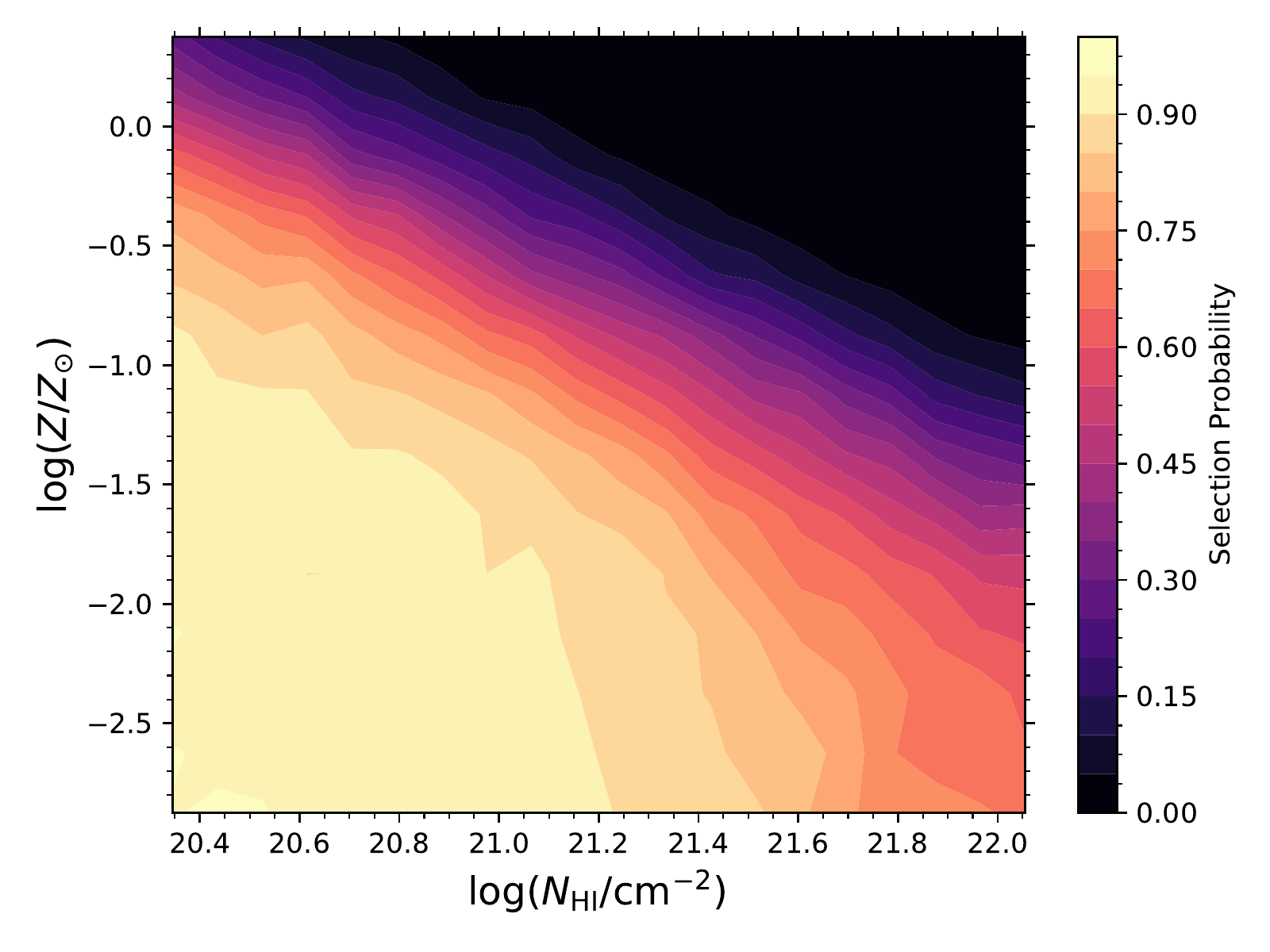}
		\includegraphics[width=0.5\textwidth]{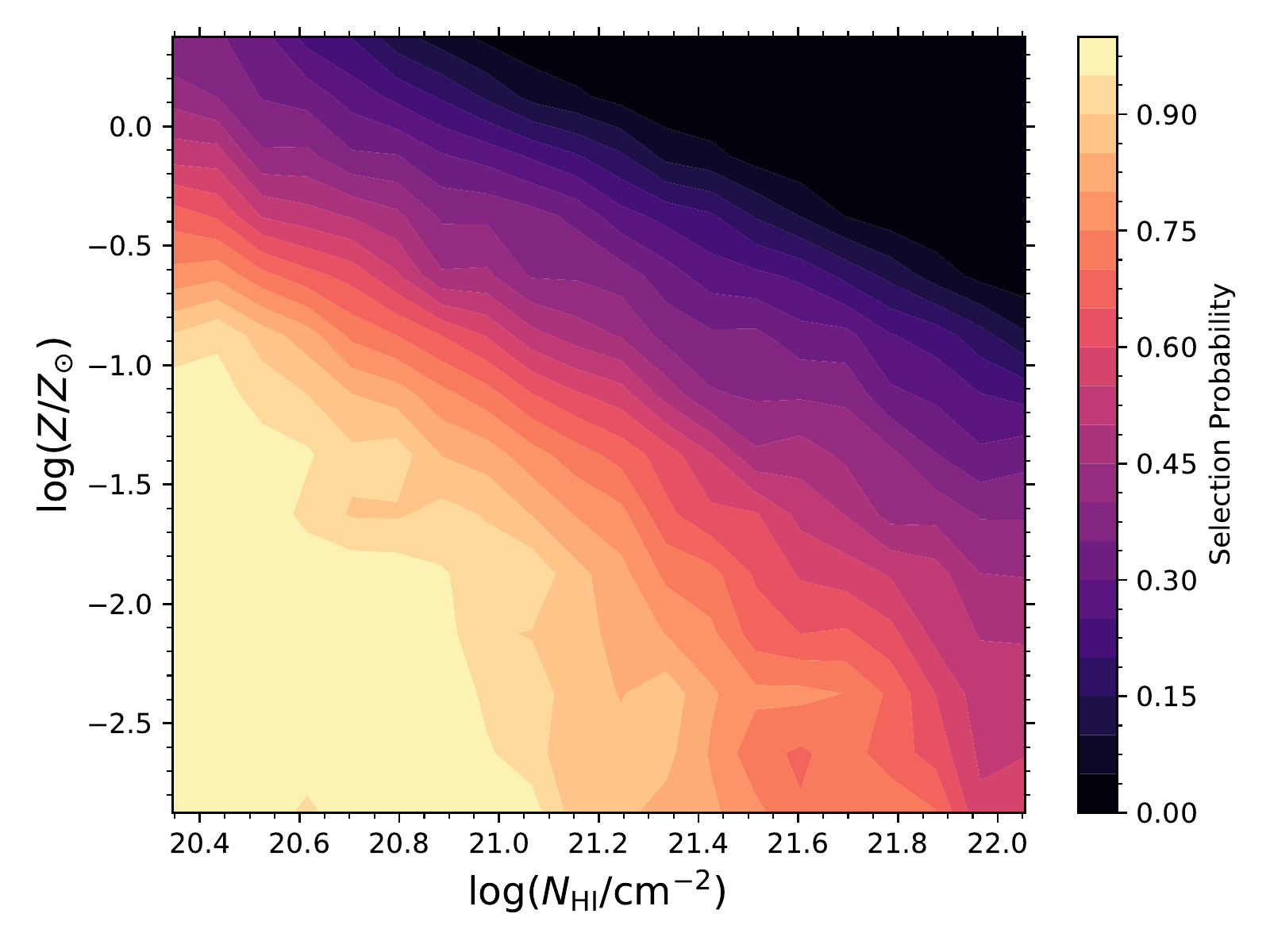}
		\caption{
			Selection probability as function of metallicity and \ion{H}{i} column density
			assuming \zqso~$=3.2$ and \zabs~$=3.0$ (top), and \zqso~$=2.5$ and \zabs~$=2.2$ (bottom). Both models assume the SMC extinction curve and a dust-to-metals ratio of $\log\kz = -21.5$.}
		\label{fig:Z_bias_SMC}
	\end{figure}

	We evaluate the selection probability at the specific redshift as a function of
	$Z$ and $\NHI$, denoted $P(Z, {\rm N_{H\, \textsc{i}}})$, in the ranges of
	$-3 < \log(Z/Z_{\odot}) < 0.5$ and $20.3 < \log({\rm N_{H\, \textsc{i}}}) < 22$,
	by uniformly drawing 200\,000 sets of $Z$ and $\NHI$.
	For each draw, we randomly assign a realization of intrinsic quasar photometry in
	the five SDSS bands for the given \zqso\ following the same method as described
	in Sect.~\ref{simulation_quasars}.
	The optical extinction is then calculated using eq.~\eqref{eq:dust}, and the quasar
	fluxes are attenuated according to the SMC extinction law applied in the absorber
	rest-frame. Here we only focus on the SMC extinction curve as this the most commonly
	observed type of extinction for DLAs (see also Sect.~\ref{discussion:ext_laws}).
	Furthermore, we apply the extinction from the DLA absorption itself.
	As before, the quasar photometry is passed through the selection algorithm and the selection
	probability in the 2-dimensional ($Z$, $\NHI$)-space is calculated.
	In Fig.~\ref{fig:Z_bias_SMC}, we show the selection probability for two models
	of \zqso~$=3.2$ and \zabs~$=3.0$, and \zqso~$=2.5$ and \zabs~$=2.2$. Both models
	are calculated assuming a value of $\log \kz=-21.5$ for the SMC extinction law.
	It is clearly seen that the selection probability drops for high metallicity and
	high \logNHI\ due to the increasing reddening moving diagonally from the bottom
	left corner to the top right corner.

\section{Dust Bias in Damped \lya\ Absorbers}
\label{dustbias}

	We are now able to use the inferred selection probability, $P(Z, \NHI)$, to infer
	the completeness in metallicity and $\NHI$ distributions of DLAs at $z\approx3$.
	Since the colour selection is very redshift dependent, we also perform a calculation
	of the completeness for lower redshift DLAs at $z\approx2$ towards quasars at
	$\zqso \approx 2.5$ where the SDSS selection algorithm is very inefficient.
	For this purpose, we assume analytic, parametric functional forms for the intrinsic
	distributions of $Z$ and $\NHI$.
	For a given set of parameters for intrinsic distributions, we apply the selection
	probability to infer the corresponding biased distributions (i.e., the distributions
	that would be observed). The parameters for the intrinsic distributions are then
	varied in order to match the observed distributions inferred from SDSS data up until DR7.
	Specifically, we use a log-normal metallicity distribution:
	\begin{equation}
		f(Z, \mu_{\textsc z}, \sigma_{\textsc z})\,
		\propto\, \exp \left(-\frac{(Z-\mu_{\textsc z})^2}{2 \sigma_{\textsc z}^2} \right)~,
	\end{equation}
	with mean and standard deviation denoted as $\mu_{\textsc z}$ and $\sigma_{\textsc z}$,
	respectively. This functional form accurately describes the observed distribution
	\citep{Rafelski2014} as well as simulations \citep{Pontzen2008, Bird2015}.

	For the $\NHI$ distribution, we use a $\Gamma$-function:
	\begin{equation}
		f(\NHI, \alpha, N_0)\,
		\propto\, \left( \frac{\NHI}{N_0} \right)^{\alpha} \exp \left( -\frac{\NHI}{N_0} \right)~,
	\end{equation}
	as argued by \citet{Pei1995} and further supported by observations
	\citep{Noterdaeme2009b, Noterdaeme2012b}.

	We assume that the two distributions are independent since simulations of DLAs in
	cosmological contexts show no significant correlation between $\NHI$ and halo-properties
	such as mass or metallicity \citep{Pontzen2008}, see also discussion in Sect.~\ref{Z-NHI_correl}.
	We can therefore construct a 2-dimensional distribution, $F(\theta)$, as the product
	of the two distributions in the ($Z$, $\NHI$) parameter space, such that:
	\begin{equation}
		F(Z, \NHI, \theta) = f(Z, \mu_{\textsc z}, \sigma_{\textsc z}) \times f(\NHI, \alpha, N_0)~,
	\end{equation}
	where $\theta$ refers to the set of intrinsic parameters
	$\theta = \{\mu_{\textsc z}, \sigma_{\textsc z}, \alpha, N_0 \}$ for the two distributions.
	
	We then construct the biased distributions:
	\begin{equation}
		f'(Z, \theta) = \int F(Z, \NHI, \theta)\, P(Z, \NHI)\, {\rm d}\NHI
	\end{equation}
	and
	\begin{equation}
		f'(\NHI, \theta) = \int F(Z, \NHI, \theta)\, P(Z, \NHI)\, {\rm d}Z~,
	\end{equation}
	which we can compare to the actual observed distributions $f_{\rm obs}(Z)$ \citep{Rafelski2012, Jorgenson2013}
	and $f_{\rm obs}(\NHI)$ \citep{Noterdaeme2009b}. Although more recent and larger
	compilations of metallicity and \ion{H}{i} column density measurements are available
	\citep[e.g.,][]{Quiret2016, Noterdaeme2012b}, we use the data by \citet{Rafelski2012},
	\citet{Jorgenson2013} and \citet{Noterdaeme2009b} as these are, to our knowledge,
	the largest, homogeneous samples selected purely from SDSS-II up until DR7
	(see Sect.~\ref{stat_samples}).
	For later data releases, the quasars were selected based on different criteria than
	the colour selection algorithm that we investigate in this work.
	We highlight that the data presented by \citet{Rafelski2012} and \citet{Jorgenson2013}
	are compiled using various elements; sulphur is used whenever possible. If sulphur is
	not available, silicon or zinc is used, and lastly, iron is used at low metallicities
	as the authors argue that dust depletion does not significantly affect the abundances
	at low metallicity \citep{Rafelski2012, Jorgenson2013}.
	For the high-redshift absorbers (\zabs~$=3.0$), we compare to metallicity measurements
	of DLAs in the redshift range $2.5 < z_{\rm abs} < 3.2$ in order to have a large enough
	sample. For the low-redshift absorbers (\zabs~$=2.2$), we use metallicity measurements
	of DLAs in the redshift range $1.8 < z_{\rm abs} < 2.5$.
 
	The best-fit intrinsic parameters, $\theta$, for the two distributions can then be found
	by $\chi^2$ minimization:
	\begin{align}
		\chi^2 &= \sum \frac{\left(\tilde f_{\rm obs}(Z) - \tilde f'(Z, \theta) \right)^2}
							{\tilde f_{\rm obs}(Z)}\\
		&+ \sum \frac{\left(\tilde f_{\rm obs}(\NHI) - \tilde f'(\NHI, \theta) \right)^2}
							{\tilde f_{\rm obs}(\NHI)}~,
	\end{align}
	where the $\tilde f'$ and $\tilde f_{\rm obs}$ refer to the normalized distributions,
	e.g., $\tilde f_{\rm obs}(Z) = f_{\rm obs}(Z) / \int {\rm d}Z\, f_{\rm obs}(Z)$.\\

	In order to make a direct comparison to the work by \citet{Pontzen2009}, we calculate
	the fractional completeness in the same quantities as they do. These are the incidence rate:
	\begin{equation}
		l_{\rm DLA} \propto \int F(Z, \NHI, \theta)\, {\rm dN_{H\, \textsc{i}}\, d}Z \equiv \phi_0 ~;
	\end{equation}
	the total mass density of neutral hydrogen in DLAs:
	\begin{equation}
		\Omega_{\rm DLA} \propto \int {\rm N_{H\, \textsc{i}}}\, F(Z, \NHI, \theta)\, {\rm dN_{H\, \textsc{i}}\, d}Z~;
	\end{equation}
	the mean metallicity of DLAs:
	\begin{equation}
		\langle Z \rangle \propto \int Z\, F(Z, \NHI, \theta)\, {\rm dN_{H\, \textsc{i}}\, d}Z / \phi_0~;
	\end{equation}
	the total mass density of metals in DLAs:
	\begin{equation}
		\Omega_Z \propto \int {\NHI}\,Z\, F(Z, \NHI, \theta)\, {\rm dN_{H\, \textsc{i}}\, d}Z~;
	\end{equation}
	and the column density weighted mean metallicity of DLAs:
	\begin{equation}
		\langle Z \rangle _{\rm N_{H\, \textsc{i}}} = \frac{\Omega_{Z}}{\Omega_{\rm DLA}}~.
	\end{equation}

	\noindent
	The fractional completeness, $C$, is then calculated for the above quantities as the ratio of the biased over the intrinsic quantity, e.g.:
	\begin{equation}
		C\,[l_{\rm DLA}] = \frac{\int F(Z, \NHI, \theta)\,  P(Z, \NHI)\, {\rm dN_{H\, \textsc{i}}\, d}Z}
							    {\int F(Z, \NHI, \theta)\, {\rm dN_{H\, \textsc{i}}\, d}Z}~.
	\end{equation}

	\noindent
	Moreover, we calculate the mean biased and intrinsic reddening as:
	\begin{equation}
		\avgEbv_{\rm obs} = \frac{\kz}{R_V \phi'_0}
		\int \NHI\,Z\, F(\theta)\, P\ {\rm d\NHI\, d}Z
	\end{equation}
	\noindent and
	\begin{equation}
		\avgEbv_{\rm int} = \frac{\kz}{R_V \phi_0} \int \NHI\,Z\, F(\theta)\, {\rm d\NHI\, d}Z ~,
	\end{equation}

	\noindent where $F(\theta)$ and $P$ are shorthand notations for $F(Z, \NHI, \theta)$
	and $P(Z, \NHI)$, respectively.

\begin{figure*}
		\centering
		\includegraphics[width=0.9\textwidth]{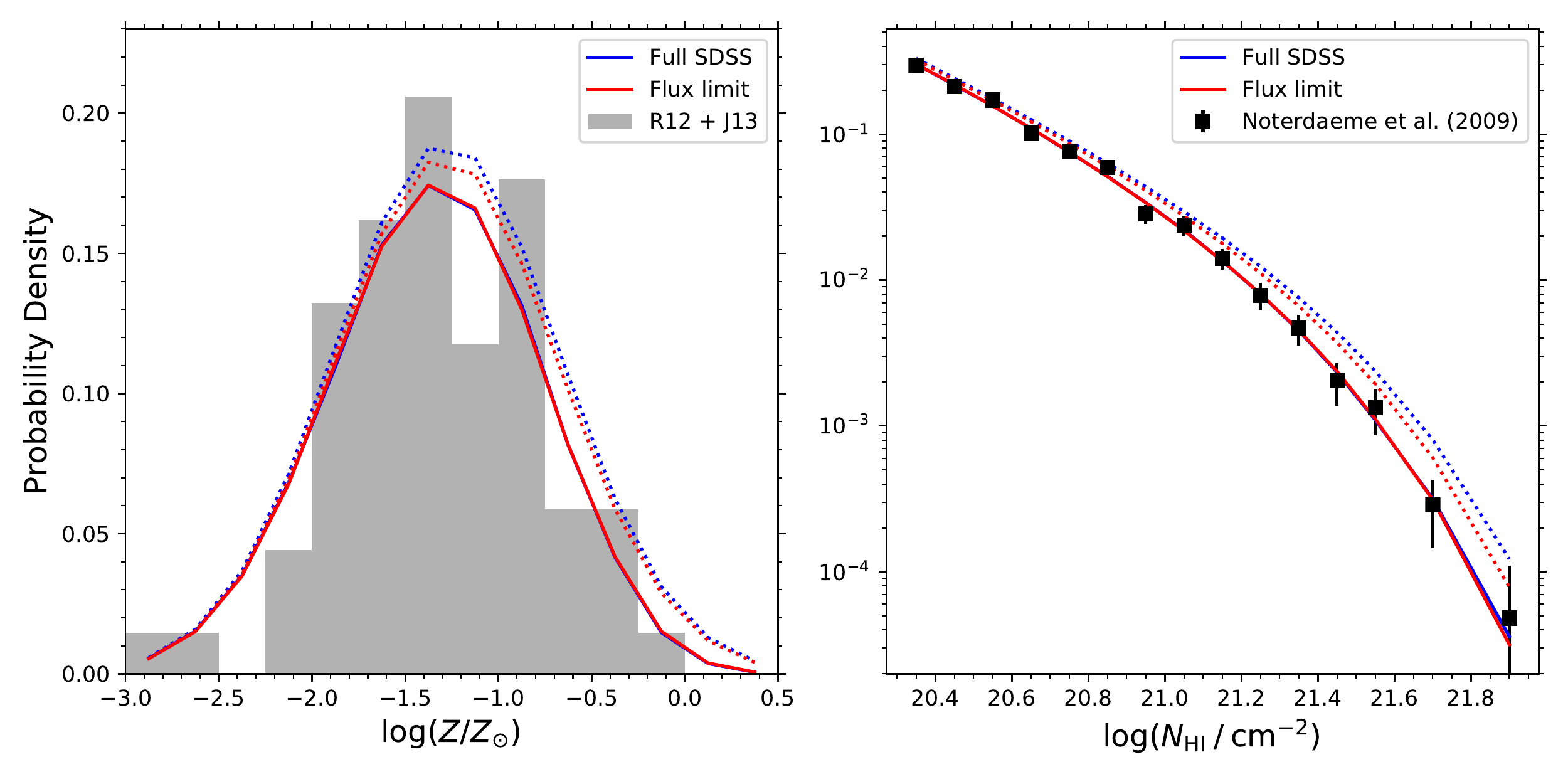}
		\caption{Metallicity and $\NHI$ distribution functions from literature.
		The dotted lines show the intrinsic, best-fit functions while the solid
		lines show the corresponding biased functions assuming either the full
		SDSS colour selection plus radio (blue) or the flux-limited selection only
		(red). The red and blue solid lines overlap making it hard to distinguish
		the two. Here we assume \zqso~$= 3.2$ and \zabs~$=3.0$ for
		SMC type dust with a single dust-to-metals ratio of $\log(\kz) = -21.5$.}
		\label{fig:intrinsic_distribution_fit}
	\end{figure*}

	In Table~\ref{tab:bias}, we summarize the results from our analysis for different
	dust-to-metals ratios, \kz, in order to match the range of probable values of the
	average observed reddening: $2 < \avgEbv < 10$~mmag \citep{Murphy2016}.
	The best-fit intrinsic distributions of $Z$ and $\NHI$ are shown in
	Fig.~\ref{fig:intrinsic_distribution_fit} as the dotted blue lines. The corresponding
	biased distributions are shown as solid lines together with the observational data,
	to which the biased distributions are fitted. The red lines result from a purely
	flux-limited selection as described below.
	In Table~\ref{tab:bias_lowz}, we summarize the results for the low-redshift
	configuration of \zqso~$=2.5$ and \zabs~$=2.2$. As the metallicity distribution
	is very sparsely sampled in this redshift range, we advise that these results
	should be interpreted with caution. Moreover, there are no direct measurements
	of the average reddening induced by DLAs in the SDSS-DR7 at this redshift.
	We therefore simply assume the same range of \kz\ as used to match the observed
	range of \avgEbv\ for the high-redshift absorbers.
	
	Lastly, we calculate the fractional completeness assuming a purely flux-limited
	selection in order to evaluate the separate contributions of the colour-selection
	and the magnitude limit and to compare our results to the previous results by
	\citet{Pontzen2009}. For the comparison to be valid, we only compare models
	with a similar dust-to-metals ratio as used by the authors.
	\citet{Pontzen2009} infer a most likely value of the dust opacity at
	$\lambda \sim 1900$~\AA\ of $\log\tau_0 \approx -21.8$, which corresponds to a
	value of \kz\ of $\log\kz \approx -21.5$\footnote{We note that Pontzen \& Pettini
	use a metallicity dependent dust model where low metallicity absorbers below a
	threshold of roughly $Z_{\rm lim} < 0.1 Z_{\odot}$ have a lower dust-to-metals
	ratio. This does not significantly alter the calculation of fractional completeness
	as this simply lowers the dust extinction at low metallicities where the completeness
	is already roughly unity. The converted value of \kz\ is thus quoted for high
	metallicities, $Z \sim Z_{\odot}$.}. For the comparison, we therefore focus
	on models with $\log\kz = -21.5$, see the red lines in
	Fig.~\ref{fig:intrinsic_distribution_fit}.
	The calculated fractional completeness for a purely flux-limited sample is given
	in Table~\ref{tab:bias_flux}. We highlight that only the high-redshift results
	should be compared to the results by \citet{Pontzen2009}, as this is the only
	redshift range considered by these authors.
	
	We find that the main contribution to the incompleteness at $z\approx 3$
	is due to the flux-limit imposed on the quasar selection.
	This is also evident from the close resemblance between the dotted blue and red lines
	in Fig.~\ref{fig:intrinsic_distribution_fit}.
	However, the fractional completeness is systematically lowered when taking
	into account the full quasar selection algorithm. The effect on the fractional
	completeness of $\Omega_Z$ is 0.05, corresponding to a decrease in completeness
	by $\sim10$\%.
	For the lower redshift model, the fractional completeness is significantly
	affected by the additional colour selection: $\Omega_{\rm DLA}$ and $\Omega_Z$
	are lowered by 0.15 (22\%) and 0.13 (37\%), respectively, when taking the full
	selection algorithm into account.

\begin{table*}
	\caption{Fractional completeness and average reddening of $z\approx 3$ DLA properties for different dust-to-metals ratios, $\kz$. \label{tab:bias}}
	\centering
	\begin{tabular}{lccccccc}
		\hline\vspace{2pt}
$\log \kz$                           &     $-21.1$             &     $-21.2$             &        $-21.3$          &        $-21.4$          &        $-21.5$          &        $-21.7$          &        $-21.9$          \\
		\hline
$C\,[\,l_{\rm DLA}\,]$               & $    0.72 \pm 0.03    $ & $    0.77 \pm 0.02    $ & $    0.81 \pm 0.02    $ & $    0.84 \pm 0.02    $ & $    0.87 \pm 0.01    $ & $    0.91 \pm 0.01    $ & $    0.93 \pm 0.01    $\\
$C\,[\,\Omega_{\rm DLA}\,]$          & $    0.61 \pm 0.03    $ & $    0.68 \pm 0.03    $ & $    0.73 \pm 0.02    $ & $    0.77 \pm 0.02    $ & $    0.81 \pm 0.02    $ & $    0.86 \pm 0.01    $ & $    0.89 \pm 0.01    $\\
$C\,[\,\langle Z \rangle\,]$         & $    0.48 \pm 0.05    $ & $    0.55 \pm 0.05    $ & $    0.62 \pm 0.05    $ & $    0.68 \pm 0.05    $ & $    0.74 \pm 0.04    $ & $    0.83 \pm 0.03    $ & $    0.89 \pm 0.02    $\\
$C\,[\,\langle Z \rangle _{\NHI}\,]$ & $    0.41 \pm 0.05    $ & $    0.48 \pm 0.05    $ & $    0.55 \pm 0.05    $ & $    0.61 \pm 0.05    $ & $    0.67 \pm 0.04    $ & $    0.76 \pm 0.04    $ & $    0.84 \pm 0.03    $\\
$C\,[\,\Omega_Z\,]$                  & $    0.25 \pm 0.04    $ & $    0.32 \pm 0.04    $ & $    0.40 \pm 0.05    $ & $    0.47 \pm 0.05    $ & $    0.54 \pm 0.04    $ & $    0.66 \pm 0.04    $ & $    0.75 \pm 0.03    $\\[2pt]
$\avgEbv_{\rm obs}$ / mmag           & $9.7\, ^{+1.0}_{-0.9} $ & $8.0\, ^{+0.7}_{-0.7} $ & $6.5\, ^{+0.6}_{-0.5} $ & $ 5.3\, ^{+0.6}_{-0.4}$ & $4.3\, ^{+0.4}_{-0.4} $ & $2.9\, ^{+0.3}_{-0.3} $ & $1.9\, ^{+0.2}_{-0.2} $\\[2pt]
$\avgEbv_{\rm int}$ / mmag           & $28.0\, ^{+5.3}_{-5.1}$ & $19.1\, ^{+3.5}_{-3.1}$ & $13.2\, ^{+2.4}_{-1.9}$ & $ 9.6\, ^{+1.7}_{-1.4}$ & $7.0\, ^{+1.1}_{-1.0} $ & $4.0\, ^{+0.6}_{-0.6} $ & $2.3\, ^{+0.4}_{-0.3} $\\[2pt]
		\hline
	\end{tabular}
\end{table*}

\begin{table}
	\caption{Fractional completeness for $z\approx2$ DLAs for different values of \kz.
			 \label{tab:bias_lowz}}
	\centering
	\begin{tabular}{lccc}
		\hline\vspace{2pt}
$\log \kz$                           &         $-21.1$           &         $-21.5$         &         $-21.9$        \\
		\hline 
$C\,[\,l_{\rm DLA}\,]$               &     $0.58 \pm 0.05$       &     $0.77 \pm 0.03$     &     $0.91 \pm 0.01$    \\  
$C\,[\,\Omega_{\rm DLA}\,]$          &     $0.49 \pm 0.05$       &     $0.68 \pm 0.03$     &     $0.84 \pm 0.02$    \\  
$C\,[\,\langle Z \rangle\,]$         &     $0.40 \pm 0.04$       &     $0.57 \pm 0.03$     &     $0.71 \pm 0.02$    \\  
$C\,[\,\langle Z \rangle _{\NHI}\,]$ &     $0.34 \pm 0.04$       &     $0.51 \pm 0.03$     &     $0.67 \pm 0.03$    \\  
$C\,[\,\Omega_Z\,]$                  &     $0.16 \pm 0.03$       &     $0.35 \pm 0.03$     &     $0.56 \pm 0.03$    \\[2pt]  
$\avgEbv_{\rm obs}$ / mmag           & $17.1\, ^{+3.5}_{-3.2}$   &  $8.0\, ^{+1.3}_{-1.2}$ & $3.3\, ^{+0.4}_{-0.4}$ \\[2pt]
$\avgEbv_{\rm int}$ / mmag           & $61.0\, ^{+12.1}_{-12.2}$ & $17.7\, ^{+3.6}_{-3.1}$ & $5.4\, ^{+0.9}_{-0.8}$ \\[2pt]
		\hline
	\end{tabular}
\end{table}

\begin{table}
	\caption{Fractional completeness for DLA statistics assuming only flux-limited selection ($i<19.1$~mag).
			 \label{tab:bias_flux}}
	\centering
	\begin{tabular}{lccc}
		\hline\vspace{2pt}
		Model               &         PP09            &   \zqso~$=3.2^{(a)}$   &    \zqso~$=2.5^{(a)}$   \\
		\hline
$C\,[\,l_{\rm DLA}\,]$               &  $0.93^{+0.07}_{-0.03}$ &     $0.90 \pm 0.01$    &     $0.88 \pm 0.02$     \\  
$C\,[\,\Omega_{\rm DLA}\,]$          &  $0.87^{+0.10}_{-0.06}$ &     $0.85 \pm 0.01$    &     $0.83 \pm 0.03$     \\  
$C\,[\,\langle Z \rangle\,]$         &  $0.75^{+0.25}_{-0.09}$ &     $0.76 \pm 0.04$    &     $0.67 \pm 0.05$     \\  
$C\,[\,\langle Z \rangle _{\NHI}\,]$ &  $0.63^{+0.19}_{-0.20}$ &     $0.69 \pm 0.04$    &     $0.59 \pm 0.06$     \\  
$C\,[\,\Omega_Z\,]$                  &  $0.56^{+0.19}_{-0.26}$ &     $0.59 \pm 0.04$    &     $0.48 \pm 0.06$     \\[2pt]  
$\avgEbv_{\rm obs}$ / mmag           &    $4^{+12}_{-2}$       & $4.4\, ^{+0.4}_{-0.4}$ &  $7.2\, ^{+0.9}_{-0.8}$ \\[2pt]
$\avgEbv_{\rm int}$ / mmag           &    $8^{+24}_{-5}$       & $6.7\, ^{+1.1}_{-0.8}$ & $13.1\, ^{+2.1}_{-2.2}$ \\[2pt]
		\hline
	\end{tabular}
	{\flushleft $^{(a)}$ The models reported here assume a fixed value of $\log\kz=-21.5$.}
\end{table}


\section{Systematic Uncertainties}
\label{systematics}

Our modelling of the dust bias in Sect.~\ref{dustbias} is based on our best-fit intrinsic
quasar model (Sect.~\ref{simulation_quasars}). It is therefore important to understand to
what degree the quasar model affects the inferred fractional completeness.
In the following section, we study the systematic uncertainty introduced by variations in
the intrinsic quasar properties. Furthermore, we assess the possibility of a non-Gaussian
metallicity distribution.

\subsection{Intrinsic Quasar Model}

	In order to test the effect of the intrinsic quasar model, we recalculate the selection
	probability $P(Z, \NHI)$ for two extreme ranges of the best-fit model parameters.
	For the first model, referred to as `l99', we offset the parameters by $-3 \sigma$
	taking into account parameter correlations, see details in Appendix~\ref{app:opt}.
	This results in a {\it bluer} intrinsic quasar shape. Similarly for the second model,
	referred to as `u99', we offset the parameters by $+3\sigma$, resulting in a {\it redder}
	intrinsic quasar model.
	We then recalculate the fractional completeness in the various properties listed in
	Table~\ref{tab:bias} as a function of \kz.
	The $3\sigma$ difference in the inferred fractional completeness is at most $0.02$,
	thus well within the statistical uncertainty. Similarly, the systematic uncertainty
	on \avgEbv\ is $\sim0.5$~mmag and $<0.1$~mmag for, respectively, the highest and
	lowest value of \kz\ considered in this work, both significantly smaller than the
	statistical uncertainty.

	The systematic uncertainty might be more significant for the lower-redshift models,
	however, due to the complexity of fitting the intrinsic quasar colour distributions
	when these are heavily affected by the selection algorithm, it is not possible to
	formally address the error in the same way as above.
	Instead, we assume the same range of parameters as used for the high-redshift model
	and offset the quasar model similarly. We then recalculate the selection probability
	and the fractional completeness.
	The systematic uncertainties are significantly stronger for the low-redshift model,
	with the quantities $l_{\rm DLA}$, $\Omega_{\rm DLA}$, and $\Omega_Z$ being mostly
	affected. For these quantities, the $1\sigma$ systematic uncertainty is of the same
	order as the statistical uncertainty yet slightly smaller ($\sim0.01$).
	
	The precise systematic uncertainties for each quantity are given in
	Table~\ref{tab:systematic} in Appendix~\ref{app:opt}.

\subsection{Non-Gaussian Metallicity Distribution}
	
	In the bias analysis above, we have assumed a log-normal distribution
	of metallicity as the intrinsic shape. However, given the strong bias
	against high-metallicity absorbers, the observed distribution might
	equally well arise from a skewed distribution where the high-metallicity
	tail has been suppressed by the dust bias.
	In the following we will investigate the effect of an intrinsically
	skewed log-normal distribution.
	We use the following parametrization of the skewed normal distribution
	\citep{OHagan1976}:
	
	\begin{equation}
		f(Z, \mu_{\textsc z}, \sigma_{\textsc z}, \zeta) =
		2 \phi \left( \frac{Z - \mu_{\textsc z}}{\sigma_{\textsc z}} \right)
		\Phi \left(\zeta \frac{Z - \mu_{\textsc z}}{\sigma_{\textsc z}} \right)~,
	\end{equation}

	\noindent
	where $\phi$ and $\Phi$ are the probability density and cumulative probability
	of the normal distribution, respectively.
	The skewness is controlled by the parameter $\zeta$, which for positive (negative)
	values will skew the distribution towards higher (lower) values than the mean.
	For $\zeta=0$, one recovers the symmetrical normal distribution.

	We then perform the same analysis as in Sect.~\ref{dustbias}, however,
	we use the skewed normal distribution of metallicity instead of the
	standard normal distribution.
	For high-redshift configuration (\zabs~$=3$), we find that the skewed
	distribution is poorly constrained: $\zeta = 1.4\pm1.3$.
	For the lower redshift model ($\zabs = 2.2$), we are not able to constrain
	$\zeta$ at all. Hence, in both cases we find no evidence for a skewed
	intrinsic distribution providing a better fit to the data.

\section{Discussion}
\label{discussion}

	In this work, we have calculated the fractional completeness of DLA statistics
	in terms of the incidence rate, the mass density of neutral hydrogen,
	the mean metallicity, the column density weighted mean metallicity,
	and the mass density of metals assuming a wide range of dust-to-metals ratios, \kz.
	In the following discussion, we focus on the high-redshift results, as these are
	better constrained by data (for a discussion of the redshift dependence,
	see Sect.~\ref{z_evolution}).
	\citet{Murphy2016} provide the latest measurement of \avgEbv\ from SDSS DR7 of
	$3.0\pm1.5$~mmag (when restricting to $i<19.1$). Moreover, the authors provide
	a review of recent measurements in the literature. The various studies reach
	different results concerning \avgEbv; nonetheless, \citet{Murphy2016} argue that
	the most probable range is $2 < \avgEbv < 10$~mmag. This range of \avgEbv\
	corresponds to the full range of \kz\ studied in this work.
	As observed from the results in Table~\ref{tab:bias}, the fractional completeness
	depends strongly on \kz.
	For the largest assumed value of $\kz=-21.1$, we infer that $\Omega_{\rm DLA}$ as
	measured in SDSS DR7 may be underestimated by 60\%, and $\Omega_Z$ by a factor of 4.
	This high value of $\log\kz$ leads to an intrinsic average reddening of
	$28$~mmag which is consistent with the limit on the unbiased
	\avgEbv\ as inferred from a purely radio-selected sample of
	$\langle E(B-V) \rangle < 40$~mmag \citep{Ellison2005}.
	However, such a large value of \kz\ is slightly at odds with the
	recent results obtained by \citet{Zafar2019} of $\log\kz = -21.4 \pm 0.1$,
	yet consistent at the 3$\sigma$ level.
	Instead, taking the \citeauthor{Murphy2016} value of $\avgEbv = 3.0\pm1.5$~mmag
	at face value, we find a most likely value of \kz\ of $\log\kz = -21.7\pm0.2$.
	This is equally at odds with the constraint by \citet{Zafar2019}.
	\citet{Murphy2016}, however, note that their measurement of $\langle E(B-V) \rangle$
	might be underestimated by up to $3$~mmag (see Sect.~6.1 and their Fig.~9).
	A correction of this size would bring their result in agreement
	with the measurement by \citet{Vladilo2008} ($\avgEbv = 6\pm2$~mmag), and instead
	favour a dust-to-metals ratio between $\log\kz = -21.4$ and $-21.3$, fully consistent
	with the result by \citet{Zafar2019}.
	
	Given the lack of strong evidence as to the true value of \avgEbv\ for DLAs
	from SDSS DR7, we instead use the constraint on \kz\ from the analysis by
	\citet{Zafar2019} of $\log\kz=-21.4\pm0.1$. Based on this value, we infer a
	fractional completeness for $\Omega_{\rm DLA}$ and $\Omega_Z$ in DLAs of
	$0.77\pm0.04$ and $0.47\pm0.09$, respectively, and we obtain an independent
	determination of the average reddening in optically-selected DLA samples of
	$\avgEbv = 5^{+2}_{-1}$~mmag. We further note that these results do not
	significantly depend on the intrinsic quasar model as the systematic
	uncertainty is much smaller than the statistical uncertainty
	(Sect.~\ref{systematics}).

	One last aspect to consider is the fraction of the population with super-solar
	metallicities, since the results must agree with constraints on the observed
	fraction of high-metallicity systems.
	For \zabs~$=3.0$ DLAs, we find that the intrinsic fraction of super-solar DLAs
	is 1-2\%.
	This is consistent with the observed fraction of galaxies with super-solar
	metallicity as probed by gamma-ray bursts (GRBs) is $4^{+5}_{-2}$~\% \citep{Kruhler2015}.
	However, as GRBs trace the galaxy population differently than DLAs
	\citep{Fynbo2008} it is not straightforward to directly compare the fraction
	of super-solar GRB-hosts to that of DLAs. Nonetheless, we argue that the inferred
	fraction of 1-2\% in an {\it unbiased} sample of DLAs is not in stark contrast
	with observations since these highly enriched DLAs would be completely missed
	in the DR7 samples due to the strong dust bias.

\subsection{Redshift Evolution of $\Omega_Z$}
\label{z_evolution}

	Due to the redshift dependent nature of the colour criteria used in SDSS DR7,
	the magnitude of the dust bias depends on redshift. Based on our calculations
	of the fractional completeness at \zabs~$=2.2$, we find that $\Omega_Z$ at this
	redshift is underestimated by a factor of $\sim$3 assuming $\log\kz=-21.5$
	as argued above. The corresponding \avgEbv\ inferred for an optically selected
	sample is $\avgEbv = 8$~mmag. While there are no direct constraints on \avgEbv\
	in this redshift range, a value of 8~mmag at \zabs~$=2.2$ is consistent with
	the results by \citet{Murphy2016}. For this value of \kz, we find a fractional
	completeness of $\Omega_{\rm DLA}$ and $\Omega_Z$ of $0.68\pm0.03$ and $0.35\pm0.03$.
	
	Taking the uncertainties on \avgEbv into account, we infer a lower estimate for
	the fractional completeness on $\Omega_{\rm DLA}$ and $\Omega_Z$ of $0.55\pm0.04$
	and $0.22\pm0.03$, respectively. These values indicate a significant redshift
	evolution of both $\Omega_{\rm DLA}$ and $\Omega_{Z}$.
	
	Such a redshift evolution of the fractional completeness of $\Omega_Z$ would
	significantly affect the measurements presented by \citet{Rafelski2014},
	and bring the redshift evolution of $\Omega_Z$ for DLAs in closer agreement
	with that inferred for star-forming galaxies (as traced by LBGs), albeit with
	a lower fraction of metals in DLAs. Still, more detailed calculations and larger
	samples of precise metallicity and reddening measurements are needed in order
	to firmly address all the redshift dependencies in the DR7 (and later) samples
	of absorbers taking into account both absorber and quasar redshifts.

	Lastly, as larger samples of DLAs are being compiled on the basis of more complex
	and varying selection criteria, the effect of a dust-bias on the metallicity
	distribution should be studied in detail. We have restricted our analysis to the
	magnitude limit of $i < 19.1$ in order to make a statistically meaningful comparison.
	However, the parent sample of DLAs include many more targets that are selected down
	to a much deeper magnitude limit ($i<20.2$ and even deeper for radio selected quasars
	and later SDSS surveys).
	One might therefore na{\"i}vely expect that the bias would be alleviated by the
	inclusion of such targets. However, since the completeness of quasars in SDSS-DR7
	for $19.1 < i < 20.2$ is only around 30\%, the direct impact on the bias is not
	easily quantified.

\subsection{Different Extinction Curves}
\label{discussion:ext_laws}

\begin{figure*}
	\centering
	\includegraphics[width=0.95\textwidth]{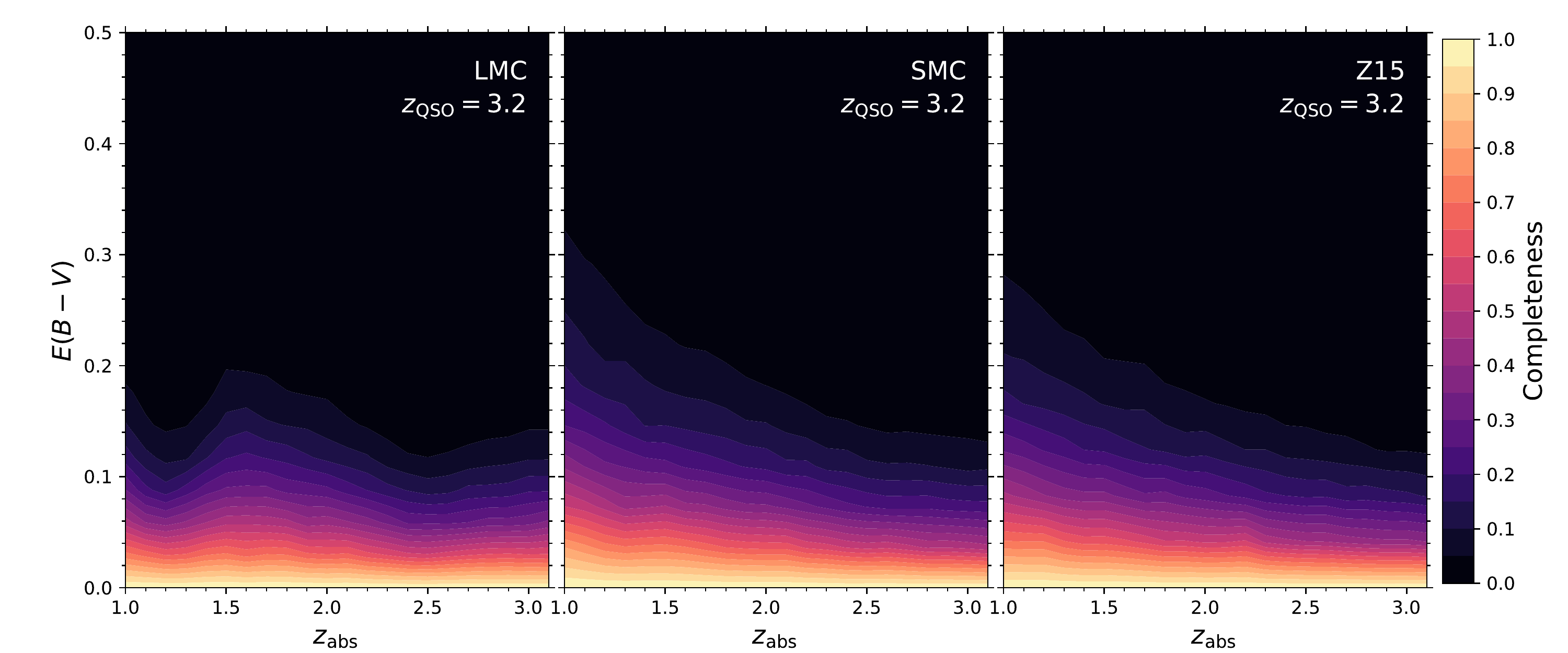}
	\caption{Completeness as a function of absorber redshift
	and optical reddening for three different reddening laws:
	LMC (left) and SMC (middle) from \citet{Gordon2003} and
	the average reddening law from \citet{Zafar2015} (right).
	The plots assume a quasar redshift of \zqso~$=3.2$ and \logNHIcm~$= 20.3$.
	\label{fig:ext_comparison}
	}
\end{figure*}

	Recently, it has been shown that dust in various environments follows a steeper
	reddening law than the SMC reddening law \citep[e.g.,][]{FM2007, Zafar2012, Zafar2015,
	Fynbo2014, Heintz2017, Noterdaeme2017}.
	In such cases, the reddening of the background quasar would be even larger than
	that of the SMC for the same \Av. We have tested the impact of inserting a steeper
	extinction curve in our simulations. For this purpose, we use the reddening law
	inferred by \citet[][hereafter Z15]{Zafar2015} and in Fig.~\ref{fig:ext_comparison}
	we show a comparison of the absorber completeness as a function of absorber redshift
	and $E(B-V)$ for three different extinction laws. We highlight that this figure,
	in contrast with similar figures in this work, shows the selection {\it completeness}
	not the selection probability, i.e., the selection probability has been corrected
	for the intrinsic quasar selection probability assuming no foreground dust.

	Based on Fig.~\ref{fig:ext_comparison}, it is seen that the smooth extinction curves
	(SMC and Z15) lead to a similar pattern of completeness as a function of \zabs\ and
	$E(B-V)$, yet with a slightly lower completeness for the steeper reddening law (Z15).
	In contrast, the characteristic 2175~\AA\ bump for the LMC-type extinction curve
	leads to a significant drop in completeness of absorbers in the redshift range
	$1 < z_{\rm abs} < 1.5$, see Fig.~\ref{fig:ext_comparison}.

	Since it is not yet possible to constrain the relative frequency of such steep
	reddening laws in DLA samples we cannot model the effects of a mixed population
	of extinction properties on the metallicity distribution. We are merely able to draw
	the qualitative conclusion that if a non-negligible part of the DLA population arises
	in galaxies with steeper reddening laws than the SMC, the bias on $\Omega_{\rm DLA}$
	and $\Omega_Z$ will be larger than we estimate in this work.
	DLAs with such steep reddening laws would preferentially be missed in the SDSS sample
	(due to the larger bias against these) and hence would not significantly alter the
	average reddening measurements that generally show good agreement with the SMC
	reddening law \citep[e.g.,][]{Vladilo2008, Murphy2016}.
	
	Similarly, there are not many quantitative estimates of the frequency of the
	2175~\AA\ bump in DLA extinction curves. The largest sample is presented by
	\citet{Zhao2016}, who identify around 400 absorbers (with $0.7 < \zabs < 2.7$)
	with LMC-type extinction curves out of roughly 40\,000 absorbers.
	However, as these are all preselected based on the \Mgii\ line, it is not clear
	what fraction of them would be DLAs.
	A proper comparison is further hampered by the fact that their full sample and
	the detection method has not yet been published. At face value though, their
	study implies that the fraction of DLAs exhibiting LMC-type extinction is less
	than 1\%. This is consistent with the \ion{C}{i}-selected sample by
	\citet{Ledoux2015} who find that roughly 1\% of DLAs are selected as \ion{C}{i}
	absorbers and roughly one third of them have LMC-type extinction.
	Given the very low frequency of LMC-type extinction in DLAs, we have focused on
	the SMC extinction law in this work.

\subsection{Correlation between $\NHI$ and $Z$?}
	\label{Z-NHI_correl}
	
	In Sect.~\ref{dustbias}, we explicitly assume that the distribution of metallicity
	and \HI\ are independent, i.e., there are no intrinsic correlation between the two
	quantities. However, if large $\NHI$ systems systematically arise in higher mass
	haloes one would expect a positive correlation between $\NHI$ and $Z$.
	If such a correlation indeed exists, then the dust bias would be enhanced, since
	there would be relatively more high-$Z$ and high-$\NHI$ absorbers for which the
	bias is stronger.
	
	Alternatively, there might be an anti-correlation between $Z$ and $\NHI$ at the
	highest column densities, since the high metallicity leads to very efficient cooling
	which in turns leads to efficient molecule formation thereby depleting the \ion{H}{i}
	into H$_2$.
	This effect would slightly alleviate the bias, since the part of the parameter space
	with high metal column (high $Z$ and high $\NHI$) would be intrinsically less populated
	and hence less affected by a dust bias.
	
	Nonetheless, no significant correlation between $\NHI$ and $Z$ is observed in
	simulations where H$_2$ formation is included at the highest $\NHI$
	\citep[e.g.,][]{Rahmati2014} nor is such a correlation reported in the observed data.
	The combination of both effects discussed above could explain the lack of any
	strong correlation between $\NHI$ and $Z$ in simulations \citep{Pontzen2008, Rahmati2014}.
	We therefore do not include any intrinsic correlations in our dust bias analysis,
	and hence assume that the two distributions are fully separable.

\subsection{DLA Identification Bias}
	\label{spectral_incompleteness}
	
	Throughout this analysis, we have made the assumption that all DLAs would be identified
	correctly if the background quasars were indeed selected as quasars based on their
	photometric properties alone. However, as DLAs are identified and searched for in the
	resulting spectra after quasar target selection, the spectral quality also impacts
	the bias. Specifically the noise level in the \lya-forest continuum is important
	since DLAs are found in this wavelength range. The largest catalog of DLAs from SDSS-II
	DR7 is based on an algorithm which only considers spectra with a continuum-to-noise (CNR)
	ratio larger than 4 in the appropriate wavelength range \citep{Noterdaeme2009b}.
	As reddened quasars with dusty foreground absorbers will have systematically
	lower CNR in the blue part of the spectrum where the DLAs are found, it is indeed possible
	that even though the quasar was correctly identified and observed, the DLA would not be
	identified in the algorithm due to the CNR being too low.
	This will therefore lead to an increase in the bias against dusty and metal-rich DLAs.
	Quantifying the additional effect would require a complete model of the spectral data
	which is beyond the scope of this paper.

\section{Summary}
\label{summary}

	In this work, we have analysed the impact of a bias against dusty absorption systems
	in SDSS DR7 on the measured cosmic mass density of neutral hydrogen and metals
	as inferred by DLAs. For this purpose, we develop a model to describe the intrinsic
	photometric properties of quasars as a function of redshift
	(Sect.~\ref{simulation_quasars}). This model is calibrated to match the observed
	colour distributions in SDSS DR7.
	Next, we implement the colour selection algorithm used by SDSS up until DR7
	\citep{Richards2002} to calculate the selection probability as a function of
	quasar redshift, details are provided in Appendix~\ref{app:algorithm}.
	
	Based on the intrinsic quasar model, we furthermore calculate the selection
	probability of quasars for various foreground absorber properties (redshift,
	visual extinction and $\NHI$).
	The selection probability of quasars and absorbers is highly redshift dependent
	and the final probability depends on both the absorber and the quasar redshift.
	In the remaining of the analysis, we focus our attention on the $z \approx 3$
	absorbers analysed by \citet{Pontzen2009} and \citet{Murphy2016}, though we
	also explore a lower redshift model.
	We then calculate the fractional completeness of the following absorption inferred
	quantities: the DLA incidence rate, $l_{\rm DLA}$, the cosmic mass density of neutral
	hydrogen in DLAs, $\Omega_{\rm DLA}$, the average metallicity, $\langle Z \rangle$,
	the hydrogen weighted average metallicity, $\langle Z \rangle _{\NHI}$,
	and the cosmic mass density of metals, $\Omega_Z$.
	
	By assuming a constant dust-to-metals ratio of $\log\kz=-21.5$ and a fixed flux-limited
	selection criterion, we are able to replicate the results by \citet{Pontzen2009}.
	However, when we include the full set of colour criteria of the SDSS DR7 selection
	algorithm including radio selection, we find that the fractional completeness is lower
	compared to a single flux-limited sample (see Tables~\ref{tab:bias} and \ref{tab:bias_flux}).
	
	The extent of the incompleteness varies with the assumed \kz. For a probable
	range in \kz\ (reproducing the observed \avgEbv), we find that $\Omega_{\rm DLA}$ and
	$\Omega_Z$ are underestimated by, respectively, $10-50$\% and $30-200$\%.
	
	At lower redshifts (\zabs~$=2.2$) assuming the same \kz, we find a fractional completeness
	of $\Omega_{\rm DLA}$ and $\Omega_Z$ of $0.68\pm0.03$ and $0.35\pm0.03$, respectively.
	However, taking all uncertainties into account, the fractional completeness
	of $\Omega_{\rm DLA}$ and $\Omega_Z$ may be as low as 0.55 and 0.22, respectively.
	We advise that these results at $z=2.2$ should be interpreted with caution as the
	metallicity distribution and the dust measurements in DLAs at this redshift are not well
	constrained, hence the resulting bias calculation is only preliminary.
	
	We conclude that while the main driver of the dust bias against metal-rich absorption
	systems at $z \approx 3$ is the flux-limit used in SDSS up until DR7, the colour
	selection adds additional incompleteness and makes the incompleteness a strong function
	of redshift. Such selection effects must be kept in mind (and preferably modelled) when
	interpreting absorber samples in any large quasar survey.
	Alternatively, the absorber statistics should be inferred using a purely unbiased
	selection such as larger radio-selected samples or Gaia-assisted selection
	\citep[][see also Geier et al. 2019, in preparation]{Heintz2018_Gaia}.

\section*{acknowledgements}
	
	We thank the anonymous referee for the thorough and constructive comments.
	JK acknowledges financial support from the Danish Council for Independent Research
	(EU-FP7 under the Marie-Curie grant agreement no. 600207) with reference
	DFF-MOBILEX--5051-00115.
	The research leading to these results has received funding from the French
	{\sl Agence Nationale de la Recherche} under grant no ANR-17-CE31-0011-01
	(project ``HIH2'' -- PI Noterdaeme).
	The Cosmic Dawn Center is funded by the DNRF.
	KEH acknowledges support by a Project Grant (162948--051) from the Icelandic Research Fund.
	The Pan-STARRS1 Surveys (PS1) and the PS1 public science archive have been made possible through
	contributions by the Institute for Astronomy, the University of Hawaii, the Pan-STARRS Project
	Office, the Max-Planck Society and its participating institutes, the Max Planck Institute for
	Astronomy, Heidelberg and the Max Planck Institute for Extraterrestrial Physics, Garching,
	The Johns Hopkins University, Durham University, the University of Edinburgh, the Queen's
	University Belfast, the Harvard-Smithsonian Center for Astrophysics, the Las Cumbres Observatory
	Global Telescope Network Incorporated, the National Central University of Taiwan, the Space
	Telescope Science Institute, the National Aeronautics and Space Administration under Grant
	No. NNX08AR22G issued through the Planetary Science Division of the NASA Science Mission
	Directorate, the National Science Foundation Grant No. AST-1238877, the University
	of Maryland, Eotvos Lorand University (ELTE), the Los Alamos National Laboratory,
	and the Gordon and Betty Moore Foundation.

\bibliographystyle{mnras}

\newpage


\appendix

\section{Optimization of Intrinsic Quasar Model}
\label{app:opt}

	We optimize the quasar model to fit more accurately the intrinsic properties of quasars in the redshift ranges used for the bias analysis, i.e., \zqso~$=3.2$ and \zqso~$=2.5$.
	For the high-z solution we do not need to take selection effects into account, since the completeness is rather high (95\%). However, for the low-z model, the model strongly depends on the colour-selection algorithm, which increases computation time beyond what is feasible for the scope of this paper. We there have to manually tweak the low-redshift model in order to match the observed properties. The details for the two models are presented below.
	
	In order to match the correlated widths of the colour distributions, we introduce an additional parameter: $w^r_{\alpha}$, the weight of the $r$-band dispersion due to intrinsic variations in the quasar spectral shape. The remaining 4 bands are all scaled using the parameter $w_{\alpha}$ as before. This can be expressed in vector notation as the 5-element vector: $\vec{w}_{\alpha} = w_{\alpha}\times\{1, 1, w^r_{\alpha}/w_{\alpha}, 1, 1\}$. Hence, the parameter $w^r_{\alpha}$ is strongly correlated with $w_{\alpha}$ by construction. The randomly assigned intrinsic variation (the second term of eq.~\eqref{eq:qso_model}) is then calculated as:
	$$
	\vec{m}_{\alpha, i} = X_i \left( \vec{w}_{\alpha} \circ \vec{\Delta}_{\alpha} \right)~,
	$$
	\noindent where ``$\,\circ\,$'' denotes the Hadamard product, i.e., component-wise multiplication.
	
	We fit the quasar model to the 4 colour distributions by minimizing the squared residual of the median, the standard deviation, the skewness (3$^{\rm rd}$ moment), and the kurtosis (4$^{\rm th}$ moment) for each of the distributions. The 4 quantities are weighted by the uncertainty on each of them which are determined by bootstrapping.
	For the high-redshift model, we optimize the 6 model parameters using a simple max-likelihood estimator. The fit is then refined using an Monte-Carlo approach in order to properly study parameter correlations.
	For this purpose, we use the Python package {\tt emcee} \citep{Foreman-Mackey2013} with 100 `walkers' initiated around the max-likelihood solution. We run the chain for 1000 iterations discarding the first 100 as `burn-in' and using `flat' priors for all parameters.
	
\begin{figure*}
	\includegraphics[width=0.98\textwidth]{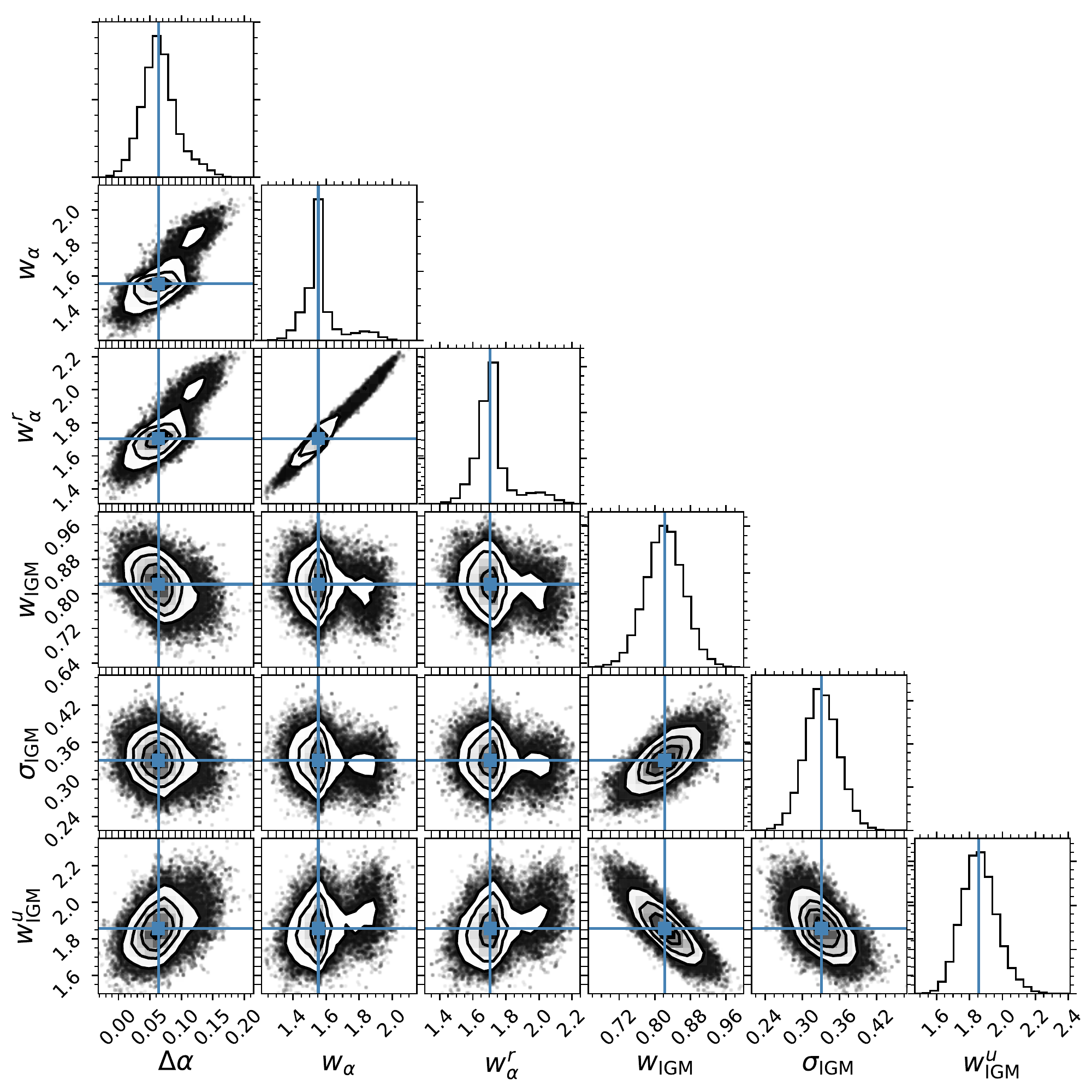}
	\caption{Posterior probability distributions and correlations
	for best-fit intrinsic quasar model for \zqso~$=3.2$.
	\label{fig:emcee}}
\end{figure*}

\begin{figure}
	\includegraphics[width=0.98\columnwidth]{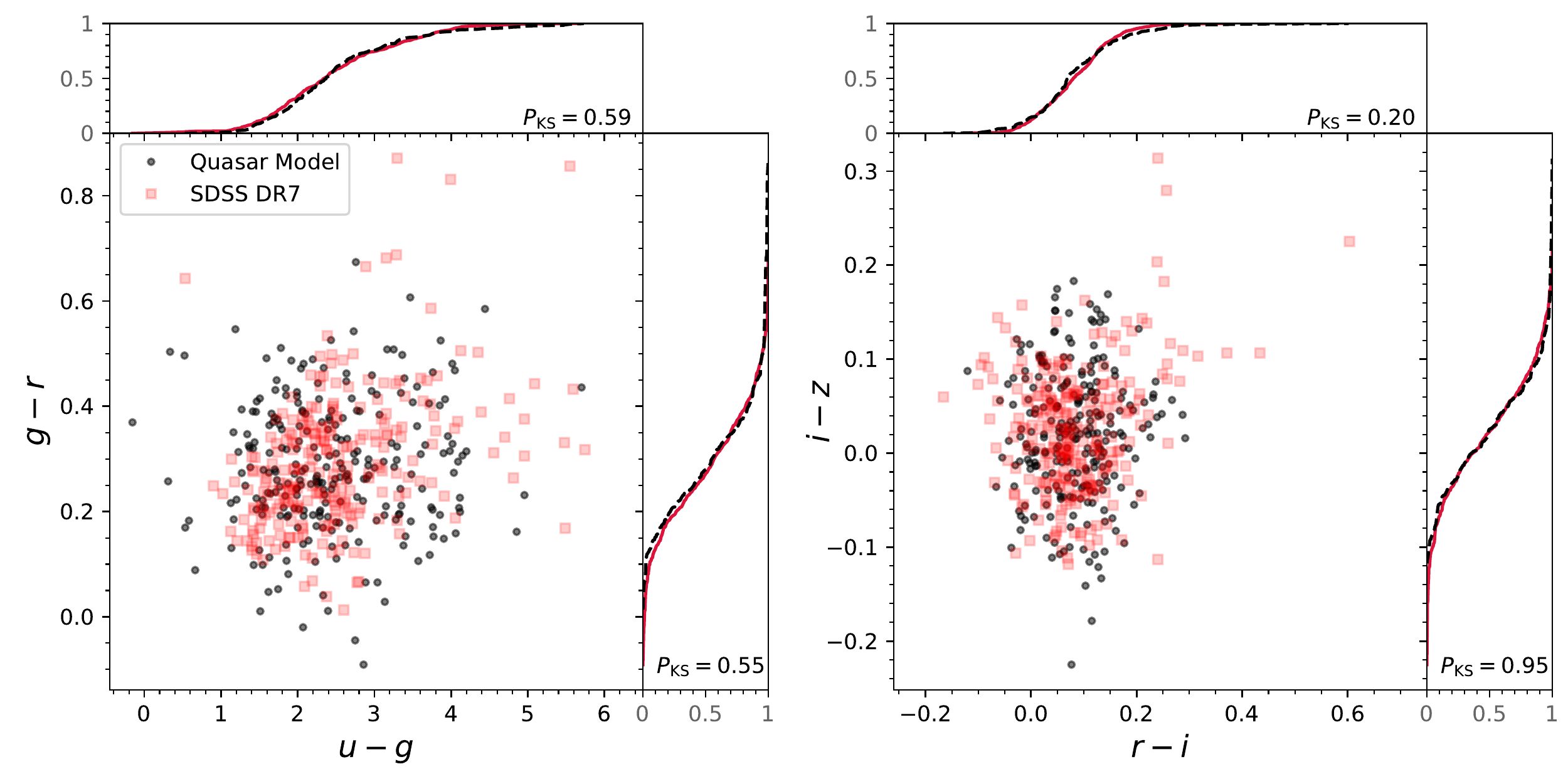}
	\caption{Colour-colour distributions for best-fit intrinsic quasar
	model for \zqso~$=3.2$. The top and right panels of each figure show the cumulative distribution of the given colour and the $p$-value derived from a 2-sample KS test.
	\label{fig:qso_model}}
\end{figure}

	The best-fit parameters are given in Table~\ref{tab:opt_pars}, and the posterior probability distributions and parameter correlations are shown in Fig.~\ref{fig:emcee}. We furthermore a comparison between the observed SDSS photometry and the model prediction in Fig.~\ref{fig:qso_model}. As indicated by the $p$-values of the 2-sample Kolmogorov--Smirnov (KS) test, all the modelled colour distributions are fully consistent with the data.
	
\begin{figure}
	\includegraphics[width=0.98\columnwidth]{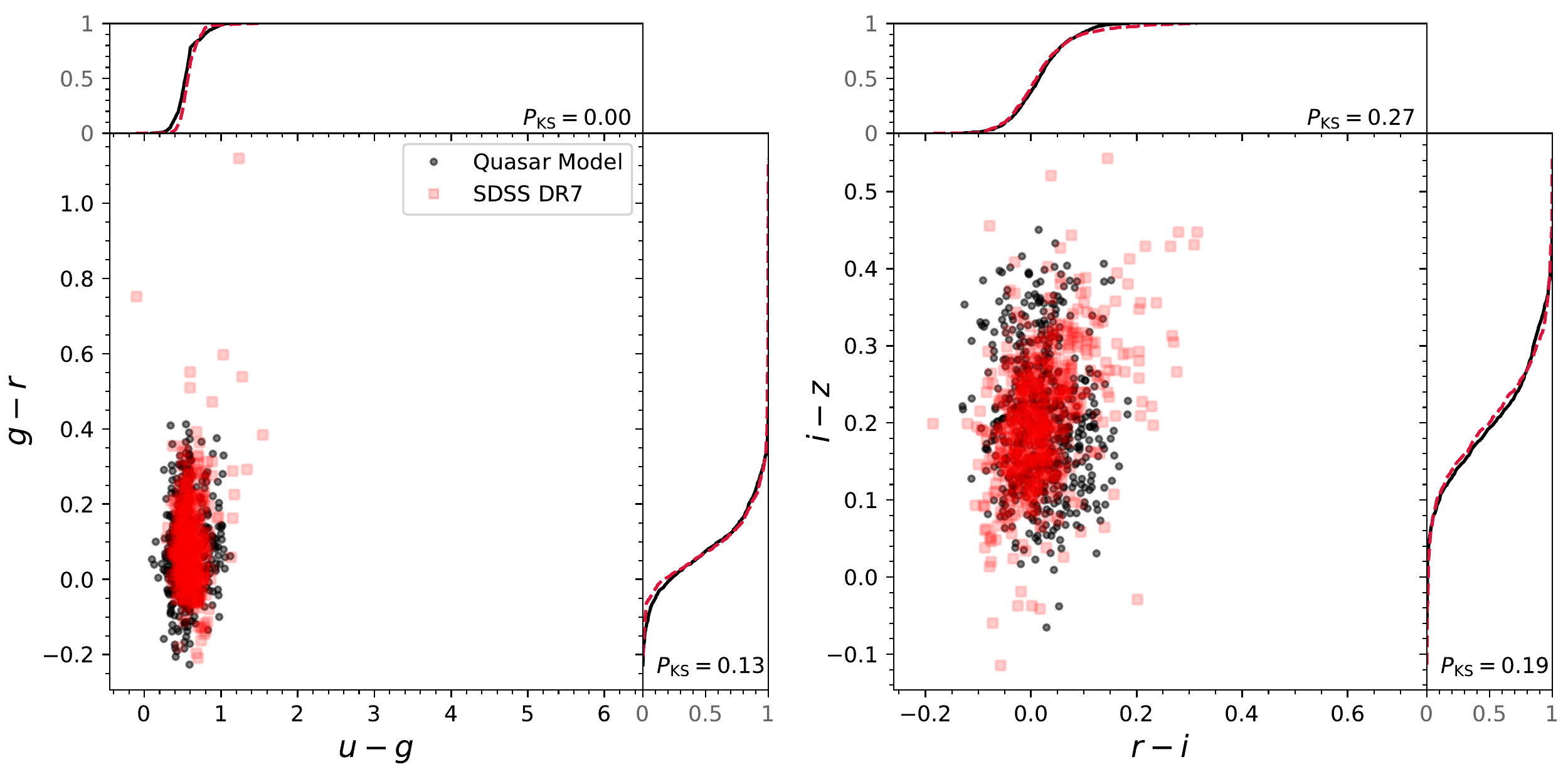}
	\caption{Colour-colour distributions for best-fit intrinsic quasar
	model for \zqso~$=2.5$. Same as Fig.~\ref{fig:qso_model}.
	\label{fig:qso_model_lowz}}
\end{figure}

	For the low-redshift model, as explained above, the colour selection algorithm strongly affects the quasar colour distributions and it is therefore not possible to do a robust Monte Carlo analysis. Moreover, we find that the 6 model parameters are not able to provide a good fit based on KS tests of the four distributions.
	Instead, we manually match the distributions by fixing the best-fit parameters and applying ad-hoc shifts to the model photometry in order to match the distributions. The applied shift per band is 0.02, 0.0, 0.088, 0.088, 0.092~mag for the $u$, $g$, $r$, $i$, and $z$ bands, respectively.
	A comparison between the observed SDSS colours and the model predictions are shown in Fig.~\ref{fig:qso_model_lowz}. The most problematic distribution to match is for the $u-g$ colour, where we are not able to fully match the shape of the distribution. However, the median is matched (by design) and the width of the distribution fits well, only the asymmetries in the distribution are not well reproduced.
	Since our main analysis is focused on the high-redshift model, we make no further attempts of improving the quasar model for this particular redshift. Moreover, due to uncertainties in the $griz$-selection (Sect.~\ref{scspy_test}), which is very important for this redshift range, the analysis at this redshift is inherently uncertain.

	\subsection{Quasar Model Systematics}
	
	In order to assess the systematic effect of varying the quasar model within the parameter uncertainties, we perform the calculation of the fractional completeness as outlined in Sect.~\ref{dustbias} for quasar models using the parameters offset by 3$\sigma$ towards bluer (l99) and redder (u99) spectral slopes. We take the parameter correlations into account when offsetting the model parameters.
	The resulting fractional completeness for the two redshift configurations are shown in Tables~\ref{tab:systematic} and \ref{tab:systematic_lowz} for a fixed value of $\log\kz=-21.5$.

\begin{table}
	\centering
	\caption{\label{tab:opt_pars}}
	\begin{tabular}{lcc}
		\hline
		Parameter       &   \zqso~$=2.5$   &   \zqso~=$3.2$  \\
		\hline
		 $\Delta\alpha$ &   $0.06\pm0.02$  &   $0.06\pm0.02$ \\
		   $w_{\alpha}$ &    $1.8\pm0.1$   &   $1.55\pm0.08$ \\
		 $w^r_{\alpha}$ &    $1.7\pm0.1$   &   $1.71\pm0.08$ \\
	 $w_{\textsc{igm}}$ &    $0.2\pm0.1$   &   $0.82\pm0.04$ \\
$\sigma_{\textsc{igm}}$ &   $0.13\pm0.03$  &   $0.33\pm0.03$ \\
   $w^u_{\textsc{igm}}$ &    $1.9\pm0.1$   &    $1.9\pm0.1$  \\
   		\hline
	\end{tabular}
\end{table}

\begin{table}
	\label{tab:systematic}
	\caption{Systematic variation in fractional completeness of
	$z\approx3$ DLA statistics and average reddening.}
	\centering
	\begin{tabular}{lcr}
		\hline
Model                                &          l99            &           u99          \\
		\hline
$C\,[\,l_{\rm DLA}\,]$               &      $0.89 \pm 0.01$    &     $0.86 \pm 0.01$    \\
$C\,[\,\Omega_{\rm DLA}\,]$          &      $0.83 \pm 0.02$    &     $0.80 \pm 0.02$    \\
$C\,[\,\langle Z \rangle\,]$         &      $0.73 \pm 0.04$    &     $0.74 \pm 0.04$    \\
$C\,[\,\langle Z \rangle _{\NHI}\,]$ &      $0.66 \pm 0.04$    &     $0.67 \pm 0.05$    \\
$C\,[\,\Omega_Z\,]$                  &      $0.55 \pm 0.05$    &     $0.53 \pm 0.05$    \\[2pt]
$\avgEbv_{\rm obs}$ / mmag           &  $4.3\, ^{+0.4}_{-0.4}$ & $4.4\, ^{+0.4}_{-0.4}$ \\[2pt]
$\avgEbv_{\rm int}$ / mmag           &  $7.0\, ^{+1.1}_{-1.2}$ & $7.1\, ^{+1.2}_{-1.1}$ \\[2pt]
		\hline
	\end{tabular}
	{\flushleft Both models are calculated for $\log\kz=-21.5$.}
\end{table}

\begin{table}
	\label{tab:systematic_lowz}
	\caption{Systematic variation in fractional completeness of
	$z\approx2$ DLA statistics and average reddening.}
	\centering
	\begin{tabular}{lcr}
		\hline
Model                       &           l99           &           u99           \\  
\hline
$C\,[\,l_{\rm DLA}\,]$               &     $0.72 \pm 0.04$     &     $0.86 \pm 0.03$     \\  
$C\,[\,\Omega_{\rm DLA}\,]$          &     $0.64 \pm 0.04$     &     $0.76 \pm 0.03$     \\  
$C\,[\,\langle Z \rangle\,]$         &     $0.53 \pm 0.03$     &     $0.58 \pm 0.03$     \\  
$C\,[\,\langle Z \rangle _{\NHI}\,]$ &     $0.47 \pm 0.03$     &     $0.52 \pm 0.03$     \\  
$C\,[\,\Omega_Z\,]$                  &     $0.30 \pm 0.03$     &     $0.39 \pm 0.03$     \\[2pt]  
$\avgEbv_{\rm obs}$ / mmag           &  $8.1\, ^{+1.5}_{-1.3}$ &  $8.1\, ^{+1.1}_{-1.3}$ \\[2pt]
$\avgEbv_{\rm int}$ / mmag           & $19.5\, ^{+3.6}_{-3.9}$ & $17.6\, ^{+2.6}_{-3.4}$ \\[2pt]
		\hline
	\end{tabular}
	{\flushleft Both models are calculated for $\log\kz=-21.5$.}
\end{table}

\section{Supporting Figures}
\label{app:figures}

\begin{figure*}
		\centering
		\includegraphics[width=0.8\textwidth]{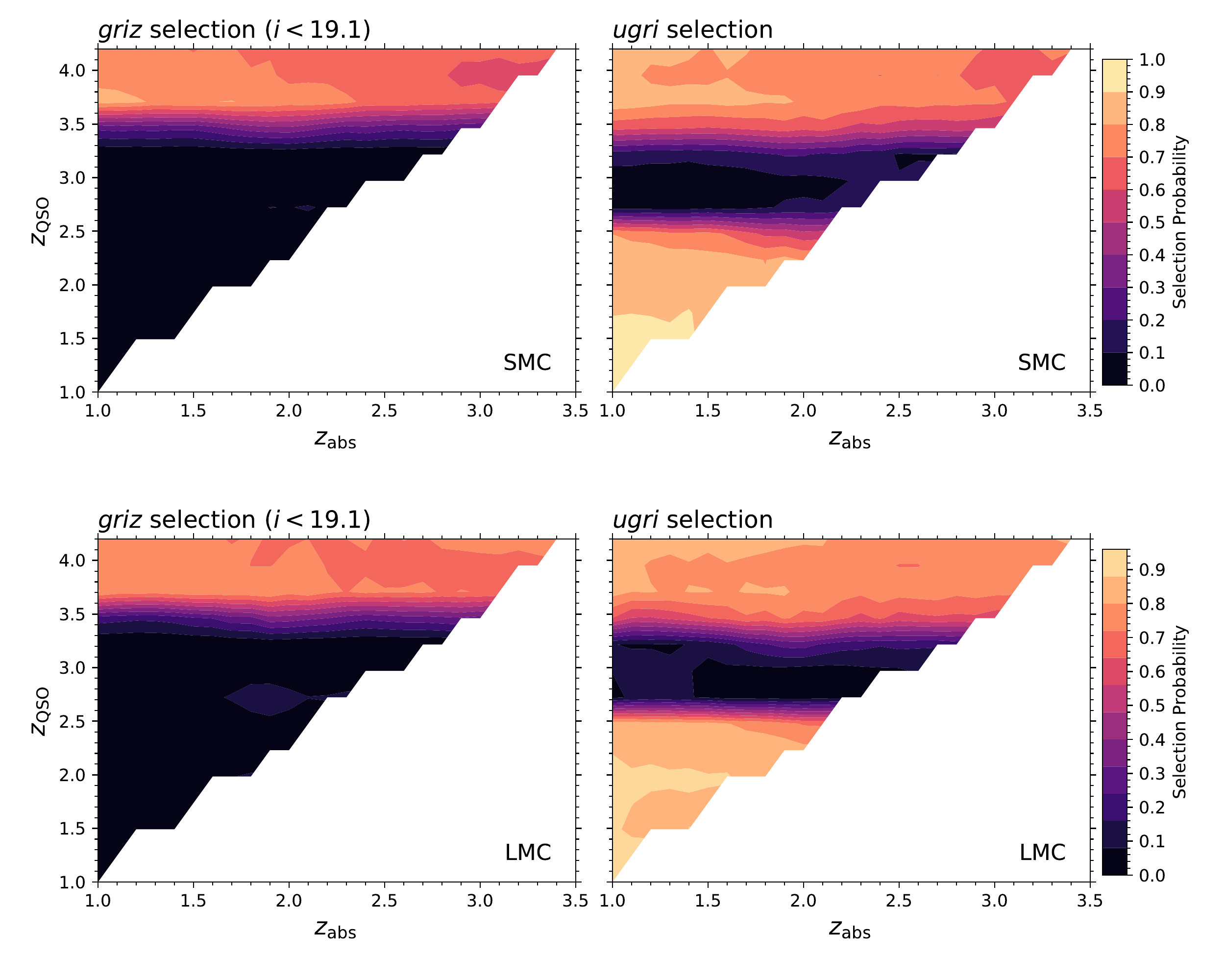}
		\includegraphics[width=0.8\textwidth]{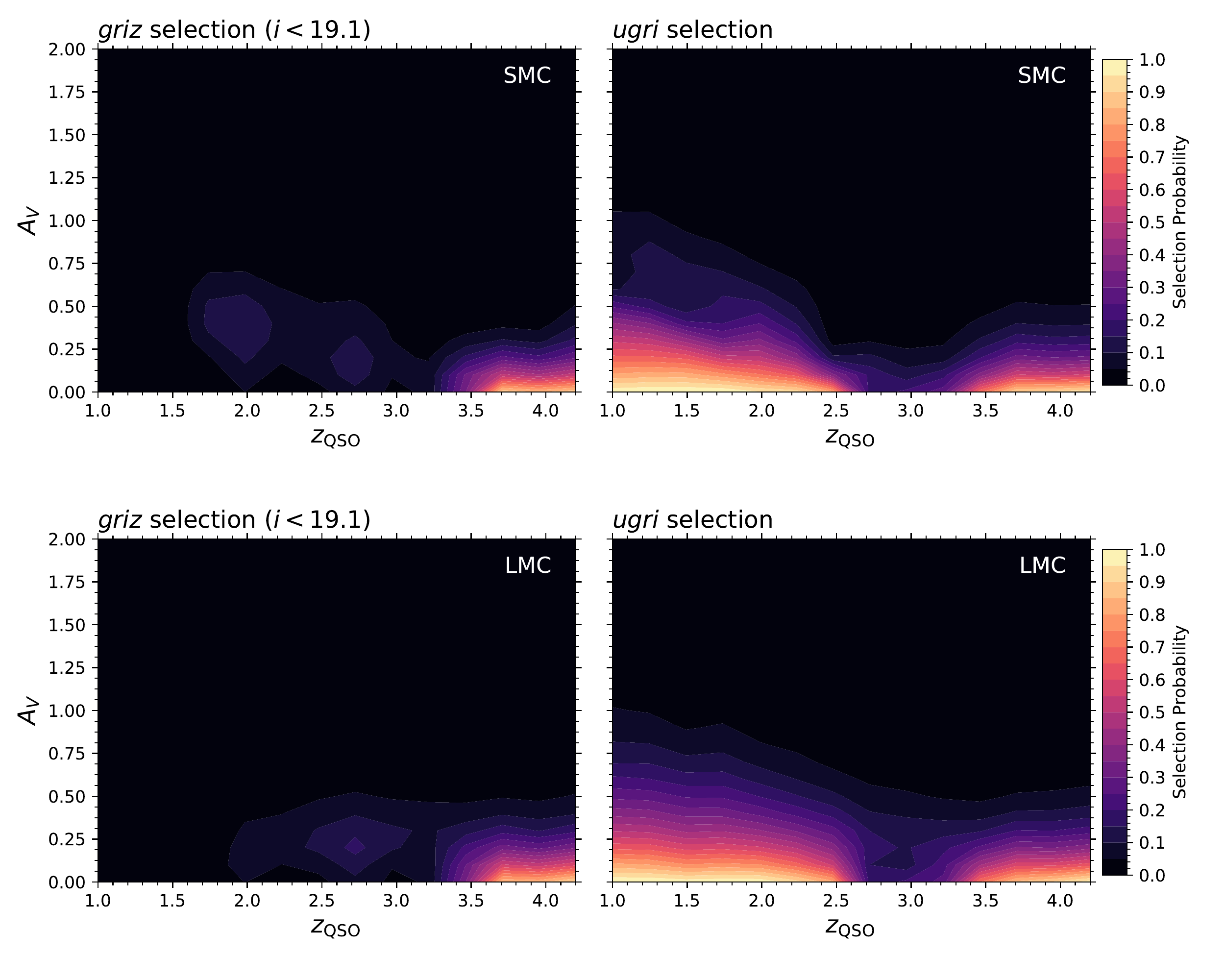}
		\caption{Selection probability as function of \zabs\ vs \zqso\ with \Av~$=0.2$~mag
		(top two rows) and \zqso vs \Av\ (bottom two rows).
		The left and right panels show the probabilities for the criteria of
		$griz$ (only for $i<19.1$) and $ugri$, respectively.
		The extinction curve used for the given model is indicated in the top right
		corner of each panel.}
		\label{fig:redshift_grid}
	\end{figure*}

\begin{figure*}
		\centering
		\includegraphics[width=0.8\textwidth]{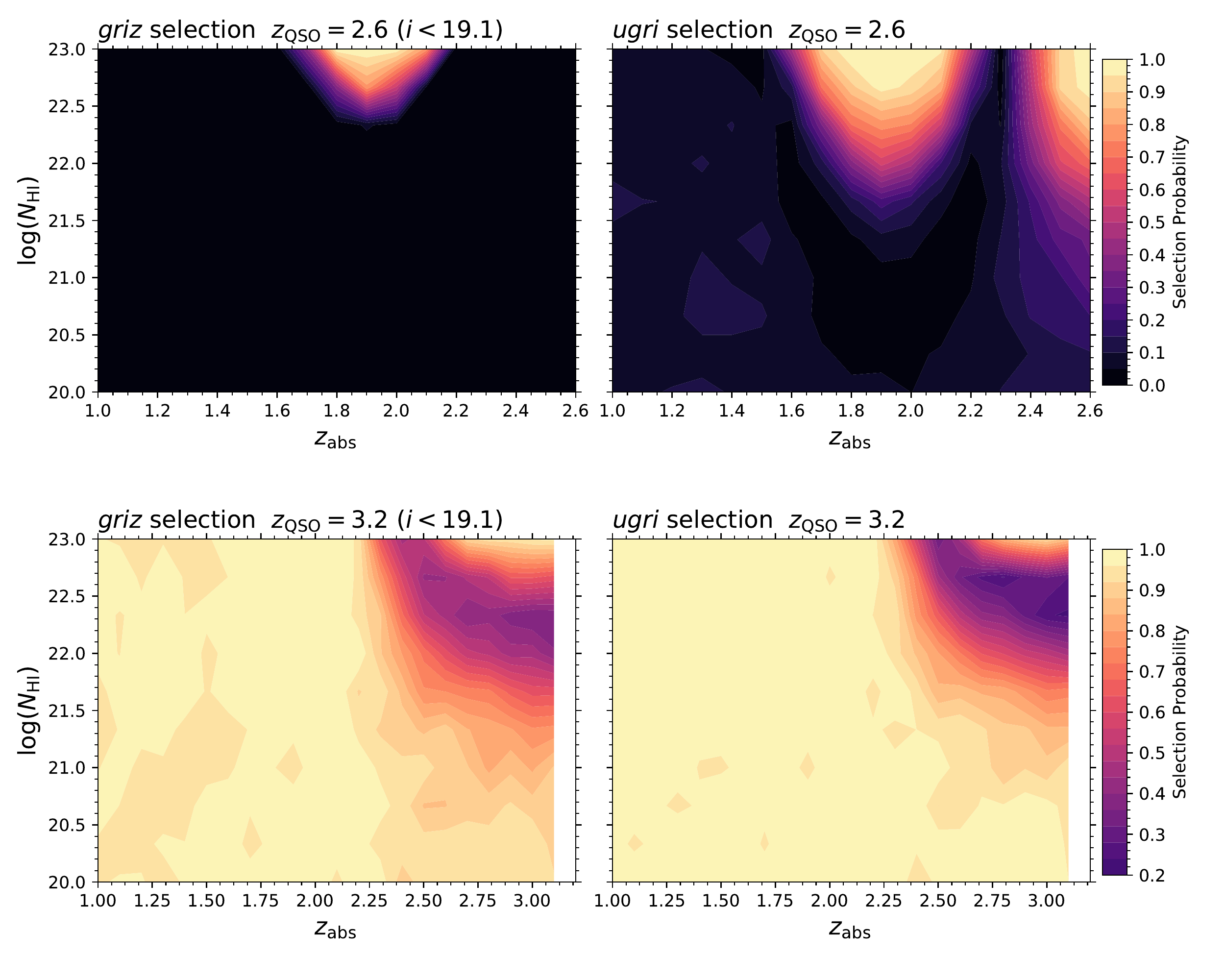}
		\caption{Selection probability as function of \zabs\ vs $\NHI$ for \Av~$=0.0$ assuming
		\zqso~$=2.6$ (top row) and \zqso~$=3.2$ (bottom row) for the SMC extinction law.
		The left and right panels show the probabilities for the criteria of
		$griz$ (only for $i<19.1$) and $ugri$, respectively.}
		\label{fig:P_NHI_zabs}
	\end{figure*}

	In Fig.~\ref{fig:redshift_grid}, we show the calculated selection probability as a
	function of \zabs\ and \zqso\ for \Av~$=0.2$~mag in the top two rows.
	In the bottom two rows of Fig.~\ref{fig:redshift_grid}, we show the selection
	probability as a function of \zqso\ and \Av\ averaged over all absorption redshifts and $\NHI$.
	In Fig.~\ref{fig:P_NHI_zabs}, we show the quasar selection probability as a function
	of $\NHI$ and \zabs\ for fixed visual extinction (\Av~$=0$) and two different quasar
	redshifts.

\section{Implementation of SDSS Colour Selection Algorithm}
\label{app:algorithm}

\subsection{Implementing the selection algorithm in Python}
	\newcommand{\mat}[1]{\mathbfss{#1}}
	\renewcommand{\vec}[1]{\mathbfit{#1}}
	\newcommand{\norm}[1]{\mathbfit{\^{#1}}}

	We follow the outline of the selection algorithm as presented by \citet{Richards2002}
	using their tabulated definition of the stellar locus in the two 3-dimensional
	colour-spaces ($ugri$ for the three colours: $u-g$, $g-r$ and $r-i$;
	and $griz$ for the three colours: $g-r$, $r-i$ and $i-z$). The coordinates in 3D
	colour space is denoted ($x$, $y$, $z$).
	The stellar locus is defined as a sequence of ellipsoids and elliptical cylinders
	centred on so-called `locus points' in the 3D colour-space.
	For each locus point, there is an associated major ($a_l$) and minor ($a_m$) axis
	of the ellipse forming the base of the cylinder, a cylinder length $k$,
	an angle $\theta$ defining the orientation of the major axis in the
	ellipse-plane\footnote{The angle is given with respect to the basis vector
	$\norm{i} = (\norm{k} \times \norm{z}) \times \norm{k}$ in the ellipse-plane.}
	denoted by coordinates ($i$, $j$), and a normal vector $\norm{k}$ perpendicular
	to the ellipse-plane indicating the direction of the ellipse along the stellar locus
	in the 3D colour-space.
	Any target with observed magnitudes in the 5-band photometric system describes a point
	in each of the two 3D colour spaces denoted by a vector $\vec{x}$.
	The first step in order to assess whether an observed point lies within the locus
	is to identify the nearest locus point to the data point $\vec{x}$. This is calculated
	by the euclidian distance in colour-space.
	
	Since a given data point in colour-space has an associated uncertainty,
	any data point is represented by an error ellipsoid determined
	by its centre ($\vec{x}$) and covariance matrix. We follow the definition
	of the covariance matrix given by \citet{Richards2002} who assume
	no correlation between individual {\it filters}. This translates to a covariance
	between two colours with the same filter occurring in both colours,
	e.g., ${\rm Cov}(g-r, r-i) = -V(r)$. An example of the covariance matrix
	for the $ugri$ colour-space is as follows:
		
		\begin{equation}
			\mat{S} = \begin{bmatrix}
				 \sigma_u^2 - \sigma_g^2 & -\sigma_g^2 & 0 \\
				-\sigma_g^2 & \sigma_g^2 - \sigma_r^2 & -\sigma_r^2 \\
				 0 & -\sigma_r^2 & \sigma_r^2 - \sigma_i^2 
			\end{bmatrix}~,
		\end{equation}
	
	\noindent
	where $\sigma$, denotes the 1~$\sigma$ uncertainty in a given filter
	
	In order to take the measurement uncertainty into account when testing whether
	a point falls inside the locus, the base-ellipse of the nearest locus-point
	cylinder is convolved with the error ellipse which arises from the projection
	of the error ellipsoid onto the ($i$, $j$) ellipse-plane.
	The data points are required to fall 4$\sigma$ away from the stellar locus.
	The covariance matrix is therefore multiplied by $N_{\sigma}^2 = 16$. 
	This projection and convolution is implemented in terms of linear algebra and
	this is the only place where we do not directly follow \citet{Richards2002}.
	However, we have tested that the two methods yield the exact same results.
	The projection is based on linear algebra equations from \citet{Pope2008}
	and is performed as follows:
	
	\begin{enumerate}
		\renewcommand{\theenumi}{\bf\Roman{enumi}}
		\setlength\itemsep{8pt}
		
		\item{Define a projection matrix, \mat{P}, which has the dimension ($2\times3$),
			  containing the two basis vectors of the locus ellipse-plane:
			  $\norm{j} = \norm{k}\times\norm{z}$
			  and $\norm{i} = \norm{j} \times \norm{k}$.}

		\item{The inverse covariance matrix of the error ellipse in the ($i$, $j$)
			  ellipse-plane is thus:
			  $\mat{A}^{-1} = \mat{G}^{-1} \mat{P} \mat{B} \mat{P}^{\intercal} \mat{G}^{-1}$,
			  where \mat{B} and \mat{G} are given by:
			  $$\mat{B} = \mat{S}^{-1} - \frac{\mat{S}^{-1}\, \norm{k}\,\norm{k}^{\intercal}\mat{S}^{-1}}
			  							 {\norm{k}^{\intercal}\mat{S}^{-1}\,\norm{k}}
			  ~~{\rm and}~~ \mat{G} = \begin{bmatrix}
			  	\norm{i}^{\intercal}\norm{i}  &  \norm{i}^{\intercal}\norm{j} \\[3pt]
			  	\norm{j}^{\intercal}\norm{i}  &  \norm{j}^{\intercal}\norm{j}
			  \end{bmatrix}~.$$}

		\item{Construct the covariance matrix defining the base of the locus
			  point cylinder in the ($i$, $j$) plane by applying a rotation to the diagonal
			  matrix, \mat{D}, by the angle $\theta$:
			  
			  $$ \mat{E} = \mat{R} \mat{D} \mat{R}^{\intercal} ~,$$
			  
			  with

			  $$\mat{D} = \begin{bmatrix}
			  	a_l^{2}  &  0 \\
			  	0  &  a_m^{2} \\
			  \end{bmatrix}
			  ~~{\rm and}~~
			  \mat{R} = \begin{bmatrix}
			  	\cos\theta  &  -\sin\theta \\
			  	\sin\theta  &  \cos\theta \\
			  \end{bmatrix}~.$$

			  }

		\item{The convolution of the error ellipse, defined by \mat{A}, and the cylinder
			  base in the ellipse-plane, defined by \mat{E}, is then simply obtained as
			  the sum of the two covariance matrices: $\mat{C} = \mat{A} + \mat{E}$.}

	\end{enumerate}

	After convolving the base ellipse of the locus point cylinder, we project the
	data point onto the ellipse-plane, i.e., from colour-space ($x$, $y$, $z$) to
	ellipse-plane ($i$, $j$).
	We then check whether the projected point, $\vec{p}_{ij}$, lies within the convolved
	cylinder defined by the covariance matrix $\mat{C}$.
	The last step is to calculate whether the point falls inside the cylinder length,
	which is similarly convolved by the projected variance along $\norm{k}$.
	
	The first locus point is enclosed by a half ellipsoid defined by the two axes of
	the cylinder ellipse together with a third axis pointing along $-\norm{k}$ of
	length $a_k$ (which is taken to be 0.2 and 0.5 for the $ugri$ and $griz$ colour-spaces,
	respectively).
	If $\vec{x}$ is closest to the first locus point, it is therefore necessary to
	test whether the data point falls within this locus ellipsoid. Again, in order
	to take the photometric error into account we convolve the base ellipse as
	specified above.
	Lastly, we convolve the third axis $a_k$ by the error by projecting the error
	ellipsoid onto the line defined by $\norm{k}$ which gives the measure $\sigma_k$.
	The length of the expanded half-ellipsoid is then given as the quadratic sum of
	$a_k$ and $\sigma_k$. We subsequently test whether $\vec{x}$ lies within the
	expanded half-ellipsoid.
	
	If a data point falls inside the stellar locus, it will not be considered any
	further as a quasar candidate. In addition to the locus-exclusion step,
	\citet{Richards2002} include a series of colour and magnitude criteria
	which we similarly implement. These are given by \citet{Richards2002} and
	we shall therefore not replicate those steps here.
	
	Since we only use the code on simulated photometry, we do not implement the
	photometric flag criteria nor do we include the handling of non-detections
	in certain bands.
	
	All the equations necessary for the projections and operations presented above
	are implemented in a Python module, {\sc scspy}, and the code is
	publicly available on GitHub\footnote{\url{https://github.com/jkrogager/scspy}}.

\subsection{Testing the selection algorithm}
\label{scspy_test}
	In order to test the algorithm, we have compiled a random set of 10\,000
	quasars from SDSS-DR7 for which photometry is available in all 5 bands.
	For the tests here, we use the dereddened photometry as is the case
	in the actual target selection algorithm.
	We have obtained the target selection information from the SDSS
	database\footnote{\url{http://skyserver.sdss.org/CasJobs/}}
	stored as the {\tt prim\_target} flag together with photometric flags
	indicating any possible errors with the photometry.
	Out of these 10\,000 quasars, 8\,417 quasars pass the photometric requirements
	imposed by \citet{Richards2002}. Before testing the colour selection algorithm,
	we include the additional systematic uncertainties \citep[][Sect.~3.4.2]{Richards2002},
	i.e., 0.0075~mag together with 15\% of the Galactic
	reddening value are added in quadrature to the photometric uncertainty.
	Of these 8\,417 quasars, only 6\,460 have been selected purely on the basis
	of colour selection, i.e., they have at least one of the primary target
	flags `QSO\_SKIRT', `QSO\_CAP' or `QSO\_HIZ'.
	When we pass the quasar photometry through our re-implementation of the
	algorithm, we recover 6\,103 objects (94\%) as quasars based on colour selection.

	We note that since it is not stated which photometric dataset was used
	for the original classification, from which the {\tt prim\_target} flag
	stems, we are not able to reproduce the classification on a one-to-one basis.
	However, on a statistical basis we are able to recover the vast majority of
	quasars.
	This is further demonstrated by the fact that 280 of 6\,101 quasars
	flagged as `QSO\_SKIRT' or `QSO\_CAP' by the SDSS $ugri$ selection have
	$i\geq 19.1$. Yet the target selection algorithm only considers targets
	with $i< 19.1$ as $ugri$ candidates \citep{Richards2002}.
	When only regarding quasars which fulfil the $i$-band magnitude cuts for the
	$ugri$ and $griz$ selection branches, we are able to recover 99\% of the quasars.\\
	
\begin{figure}
		\centering
		\includegraphics[width=0.45\textwidth]{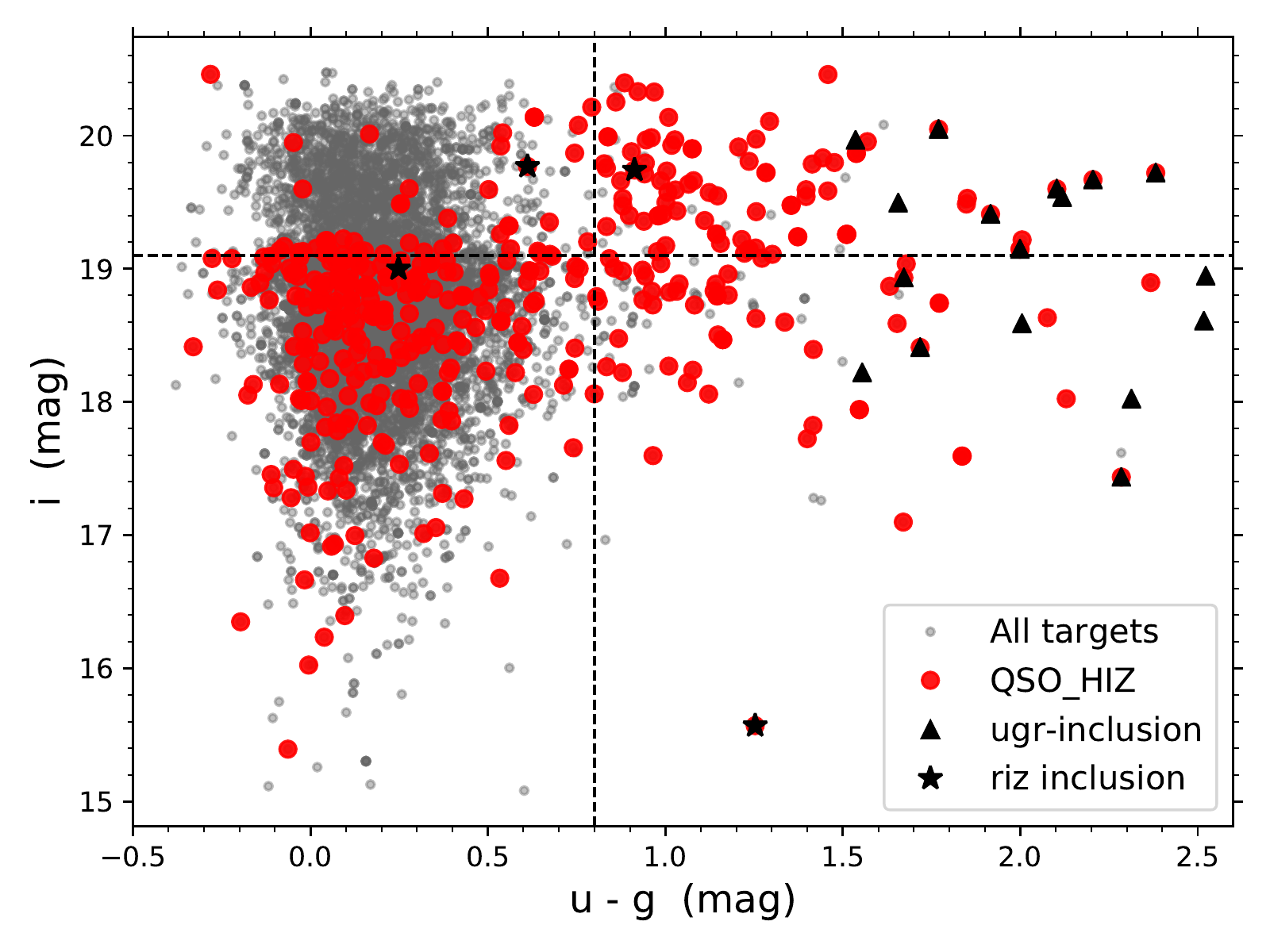}
		\caption{Colour-magnitude diagram of quasars from the DR7 test dataset
		with $g-r<1$. The targets selected as `QSO\_HIZ' from the $griz$-branch
		are highlighted as large, red dots. Targets selected from one of the
		inclusion regions are shown as black stars and triangles.
		The dotted lines mark the limits ($u-g \geq 0.8$ and $i\geq 19.1$; see text) used
		in the criteria for the low-$z$ rejection \citep[see eq. 1 of][]{Richards2002}.}
		\label{fig:lowz_rej_ugi}
	\end{figure}

	In testing the algorithm, we discovered a mistake in eq. 1 from the work by
	\citet{Richards2002} describing the low-redshift rejection criteria in the $griz$
	selection branch. When running the algorithm as specified in \citet{Richards2002},
	we find that 9 times too many targets are classified as `QSO\_HIZ'.
	Investigating the colours and magnitudes of the `QSO\_HIZ' candidates
	in the $u-g$ vs. $i$-band magnitude plane reveals that the rejection criteria
	of eq. 1 do not match the distribution of the `QSO\_HIZ' quasars in the test dataset
	(see Fig.~\ref{fig:lowz_rej_ugi}).
	Eq. 1 of \citet{Richards2002} specify that quasars with $g-r < 1$, $u-g \geq 0.8$,
	and $i\geq 19.1$ or $u-g < 2.5$ are {\it not} targeted as `QSO\_HIZ'.
	However, the distribution of $u-g$ vs $i$-band magnitude for `QSO\_HIZ'
	targets in the test data with $g-r < 1$ does not follow the criteria laid out in eq. 1.
	According to the criteria in eq. 1, there should be no targets in the range $0.8 \leq u-g < 2.5$
	irrespective of their $i$-band magnitude, i.e., in Fig.~\ref{fig:lowz_rej_ugi} there
	should be no targets selected as `QSO\_HIZ' (large red dots), only targets selected
	from inclusion regions (black stars and triangles) should be included.
	This is, however, clearly not the case.
	Since the criteria in eq. 1 were designed to exclude low-$z$ quasars fainter than $i\geq 19.1$
	we find it more plausible that the criteria should read $g-r < 1$, $u-g \leq 0.8$, and $i\geq 19.1$.
	This set of criteria would indeed discard the upper-left corner of the colour-magnitude diagram
	where the number of `QSO\_HIZ' targets is very low.

	Nevertheless, we have not been able to recover the actual criteria that were used in the
	classification of the SDSS target selection (Richards, private communication).
	Instead, we have come up with an approach that closely recovers the classification
	of the test data described above. However, we caution that the $griz$ selection
	from our re-implemented algorithm does not provide fully robust results.
	
	This error does not significantly affect our main results,
	since the magnitude limited ($i<19.1$) sample is selected almost entirely
	from the $ugri$ branch (only 2\% are selected purely from the $griz$ branch
	for our high-redshift modelling, i.e., \zqso~$=3.2$).
	
	Similarly, the analysis by \citet{Murphy2016} is affected by the error in the
	paper describing the target selection.

\end{document}